\documentclass[lettersize,journal]{IEEEtran}
\usepackage{amsmath,amsfonts}
\usepackage{algorithmic}
\usepackage{algorithm}
\usepackage{array}
\usepackage[caption=false,font=normalsize,labelfont=sf,textfont=sf]{subfig}
\usepackage{textcomp}
\usepackage{stfloats}
\usepackage{url}
\usepackage{verbatim}
\usepackage{graphicx}
\usepackage{cite}
\usepackage{amssymb}

\begin{document}

\title{Incentive Mechanism and Path Planning for UAV Hitching over Traffic Networks}

\author{Ziyi~Lu, Na~Yu  and~Xuehe~Wang,~\IEEEmembership{Member,~IEEE}
\thanks{Z. Lu, N. Yu and X. Wang (corresponding author) are with the School of Artificial Intelligence, Sun Yat-sen University, Zhuhai 519082, China (E-mail: Luzy6@mail2.sysu.edu.cn, yuna25@mail2.sysu.edu.cn, wangxuehe@mail.sysu.edu.cn). X. Wang is also with the Guangdong Key Laboratory of Big Data Analysis and Processing, Guangzhou 510006, China.}
}



\maketitle

\begin{abstract}
Package delivery via the UAVs is a promising transport mode to provide efficient and green logistic services, especially in urban areas or complicated topography. However, the energy storage limit of the UAV makes it difficult to perform long-distance delivery tasks. In this paper, we propose a novel multimodal logistics framework, in which the UAVs can call on ground vehicles to provide hitch services to save their own energy and extend their delivery distance. This multimodal logistics framework is formulated as a two-stage model to jointly consider the incentive mechanism design for ground vehicles and path planning for UAVs. In Stage I, to deal with the motivations for ground vehicles to assist UAV delivery, a dynamic pricing scheme is proposed to best balance the vehicle response time and payments to ground vehicles. It shows that a higher price should be decided if the vehicle response time is long to encourage more vehicles to offer a ride. In Stage II, the task allocation and path planning of the UAVs over traffic network is studied based on the vehicle response time obtained in Stage I. To address pathfinding with restrictions and the performance degradation of the pathfinding algorithm due to the rising number of conflicts in multi-agent pathfinding, we propose the suboptimal conflict avoidance-based path search (CABPS) algorithm, which has polynomial time complexity. Finally, we validate our results via simulations. 
It is shown that our approach is able to increase the success rate of UAV package delivery. Moreover, we estimate the delivery time of the UAV in a pessimistic case, it is still twice as fast as the delivery time of the ground vehicle only.
\end{abstract}


\section{Introduction}

\subsection{Background}
In recent years, the continuous expansion of the e-commerce market has promoted the rapid development of the express logistics industry \cite{b1}. Especially during the epidemic, people's living habits have changed dramatically in order to maintain social distance. New logistics service models have emerged in order to avoid face-to-face contact, such as non-contact delivery \cite{b2}. However, the rapid development of urban logistics has also brought a series of problems, such as traffic congestion, air pollution, low logistics efficiency \cite{b3}. Due to its agility and mobility, the delivery services enabled by the Unmanned Aerial Vehicle (UAV) are not restricted by terrain and traffic conditions, which can freely pass through urban region during rush hours to provide logistics services flexibly and efficiently. Many companies all over the world have launched commercial UAV delivery services, such as Amazon, Google, and UPS in the United States, DHL in Germany, and JD, SF-express in China \cite{b4}. However, due to the limitation of UAV energy capacity, current UAV logistics services are limited to short-distance delivery \cite{b5}. Therefore, how to expand the range of UAV delivery services is an urgent problem to be solved.

With the development of new technologies such as 5G, Internet of Things (IoT), and Artificial Intelligence (AI), crowdsourcing, as a mode of using large-scale network users to assist in specific tasks, has received more and more attention from different application platforms, and has been widely used in various scenarios, such as food delivery riders, car sharing, and data cleaning, etc \cite{b6}. Inspired by the crowdsourcing model, the ground vehicles such as trucks can be utilized to offer riding for the UAVs, which greatly saves the battery consumption of UAVs and extend their delivery distance. Amazon has proposed a UAV-truck riding strategy to allow UAVs to land on transportation vehicles from different shipping companies for temporary transport, by making an agreement with the owner of the transportation vehicles \cite{b7}. In addition, the technology of docking UAVs on stationary or high-speed moving vehicles is relatively mature \cite{b8,b9}, which lays the foundation for the development of air-ground collaborative multimodal logistics system. 

However, this new multimodal logistics system based on crowdsourcing faces the following challenges:
\begin{itemize}
    \item \emph{Incentives for ground vehicles' cooperation.} The provision of hitch services by ground vehicles is the basis for the implementation of multimodal logistics systems. However, the ground vehicles incur hitching costs (e.g., fuel consumption) when offer riding service, and they should be rewarded and well motivated to assist UAV delivery. Moreover, the ground vehicles are heterogeneous in nature. They will randomly arrive the targeting interchange point and their private costs for offering riding service are different and unknown. Thus, the pricing design under the incomplete vehicle information is challenging. In addition, a higher monetary reward leads to a shorter vehicle response time (i.e., UAV waiting time), which improves the delivery efficiency of multimodal logistics system. For sustainable management of the multimodal logistics system, the pricing strategy should be dynamic to best balance the payment to ground vehicles and their response time. 
    \item \emph{Crowdsourcing-based multimodal logistics system modeling under time-varying traffic networks and UAV energy constraint.} By integrating the UAVs with the ground vehicles, the UAVs should decide where to hitch on ground vehicles and whether hopping between different vehicles to complete one delivery task, which creates large number of interchange points on the basis of the road network for path planning. Moreover, due to the mobility of ground vehicles and the instability of traffic conditions, the modeling of the integrated system is challenging.
    \item \emph{Design of fast and efficient task allocation and path planning algorithms for the multimodal logistics system.} The ground vehicles' arrivals and response time in the vicinity of the UAVs at a certain time are unknown. In addition, the path planning of the UAVs should be carefully designed to prevent conflicts between UAVs, such as multiple UAVs landing at the same vehicle at the same time. Such combinatorial optimization problems are often NP-hard problems and challenging to solve.
 
\end{itemize}

\subsection{Main Contributions}
To tackle the above challenges, a two-stage model is proposed to jointly analyze the incentive mechanism design for ground vehicles and path planning for UAVs. We summarize our main contributions as follows:
\begin{itemize}
    \item \emph{Crowdsourcing-based multimodal logistics system modeling.} To our best knowledge, this paper is the first work studying the crowdsourcing-based multimodal logistics system, in which the ground vehicles are invited to offer hitching services for UAVs. Such multimodal logistics system effectively extends the range and efficiency of UAV deliveries. We formulate this system as a two-stage model to jointly study the incentive mechanism design for ground vehicles and path finding for UAVs. In Stage I, a dynamic pricing scheme is proposed to motivate the ground vehicles to assist UAV delivery under the incomplete information about the ground vehicles' random arrivals and private hitching costs. In Stage II, the task allocation and path planning of the UAVs over traffic network by considering the UAV energy constraint and the conflicts between UAVs. 
    \item \emph{Dynamic pricing for ground vehicle under incomplete information.} To balance the payment to ground vehicles and their response time for efficient delivery, a optimal dynamic pricing scheme is proposed under incomplete information case regarding ground vehicles’ random arrivals and private hitching costs. We prove that if the vehicle response time is long, a high price offer should be decided to encourage the ground vehicles to assist delivery. According to the proposed dynamic pricing, the vehicle response time for each interchange station can be predicted and utilized for the path planning in Stage II.
    
    \item \emph{Algorithm Design for fast and efficient task allocation and path planning.} To reduce the computation complexity, the deployment of UAV delivery tasks is split into two layers. In task allocation layer, we propose a near-optimal algorithm with polynomial computation complexity to generate the delivery sequence. In path planning layer, to avoid the computational burden due to the order-of-magnitude increase in conflicts, we propose a conflict avoidance-based path search algorithm and prove that the bounded suboptimal paths for all UAVs can be obtained in polynomial time.
    
\end{itemize}


\subsection{Related work}
The access path planning problem for UAVs can be viewed as the Traveling Salesman Problem (TSP) \cite{b11, b12}. This type of combinatorial optimization problem is a classical NP-hard problem. Once the order of package delivery for all UAVs is found, the system begins to find the optimal delivery path for each UAV. Classical algorithms such as $\text{A}*$ or Dijkstra for solving the shortest path are often used to solve the optimal UAV trajectory problem \cite{b13,b14}. In addition, control theory is also utilized for modeling and solving UAV path planning problems \cite{b15,b16}. The constraints of UAV energy storage and conflicts between UAVs make the UAV path planning problem more challenging, which is NP-hard to get the optimal solution. Some methods such as Conflict-based search (CBS) \cite{b17}, have been shown to be efficient methods to solve this Multi-Agent Path Finding (MAPF) problem.

Utilizing the ground vehicles to extend the UAVs' delivery range has attracted wide attentions from both academia and industry. The researchers in Stanford University have shown that combining UAVs with the existing transit network can increase the delivery coverage of UAVs by up to $360\%$ \cite{b18}. However, using buses for UAV delivery lacks fault tolerance. For example, the bus schedules may not be accurate. If the UAV happens to miss the arrival time of the current bus, the UAV needs to wait longer for the next bus to arrive, or has to change its route, which may 
cause the delivery task to fail. In the crowdsourcing-based system considered in this paper, the request is sent to the ground vehicle at the time when the UAV needs the hitching service and any passing-by ground vehicle along the UAV's routes can be the UAV carrier, so there is no case of missing a vehicle. E-commerce giant Amazon has proposed in its patent that commercial vans from different operators can be used to assist UAVs in delivery services \cite{b7}. However, this is also based on the pre-signed agreement with the van operators. The number of available commercial vans is limited and cannot cover the entire transportation network, which will greatly reduce the efficiency of UAV delivery. In this paper, the ground vehicles (such as trucks, buses) with wide coverage are added to the logistics network based on crowdsourcing technology, which helps keep the waiting time for UAVs to a minimum. 

The rest of this paper is organized as follows. The system model is given in Section \ref{sec_systemmodel}, in which the multimodal logistics system is formulated as two stages. In Section \ref{sec3}, the incentive mechanism for the crowdsourcing-based hitching services is designed in Stage I. Sections \ref{sec4} and \ref{sec5} discuss the task allocation and multi-agent path finding for the UAVs in Stage II. Section \ref{sec6} validates our results via simulations. Section \ref{sec_conclusion} concludes this paper.

\section{System Model}\label{sec_systemmodel}

In this paper, we aim to solve the UAV's limited flight range problem caused by its energy constraint in the process of delivery. We propose an efficient multimodal logistics system to improve the delivery range of UAVs, in which a crowdsourcing mechanism is implemented by inviting the ground vehicles (trucks, buses, etc.) to offer the hitch service for UAVs.

\subsection{Problem}
Consider a logistics system with $N$ UAVs performing delivery tasks to $M$ known delivery addresses from $K$ depots. Denote the set of $M$ delivery addresses as $V_G=\{g_1,\dots,g_M\}\subset\mathbb{R}^2$ and the set of $K$ depot locations as $V_D=\{d_1,\dots,d_K\}\subset\mathbb{R}^2$. We assume that any package can be dispatched from any depot and the depot can also charge or change the battery of the UAV quickly. 
When a UAV delivers a package from a depot to a delivery location, the UAV can call a vehicle on the ground to provide a hitch service, which significantly extends the UAV's delivery range. In order to facilitate management, we specify that the UAV can stop and wait for ground vehicles only at specific interchange points equipped with surveillance cameras for safety's sake. We generate $L$ potential interchange routes on the traffic network starting and ending at certain interchange points, where the UAV can wait for ground vehicles while performing delivery tasks. The set of interchange points is denoted as $V_I=\{i_1,\dots,i_{L^{'}}\}\subset\mathbb{R}^2$ ($L^{'}\leq L$).

In the crowdsourcing-based multimodal logistics system, the goal is to (i) design the optimal dynamic pricing to deal with the tradeoff between the payment to ground vehicles and UAVs' waiting time; (ii) find the delivery path of each UAV to deliver all packages in a reasonable combination of UAVs' own flight path and transit routes on the traffic network, while minimizing the maximum delivery time for each UAV without exceeding the storage capacity limit of the UAV.

\begin{figure}[t]
\centerline{\includegraphics[width=0.5\textwidth]{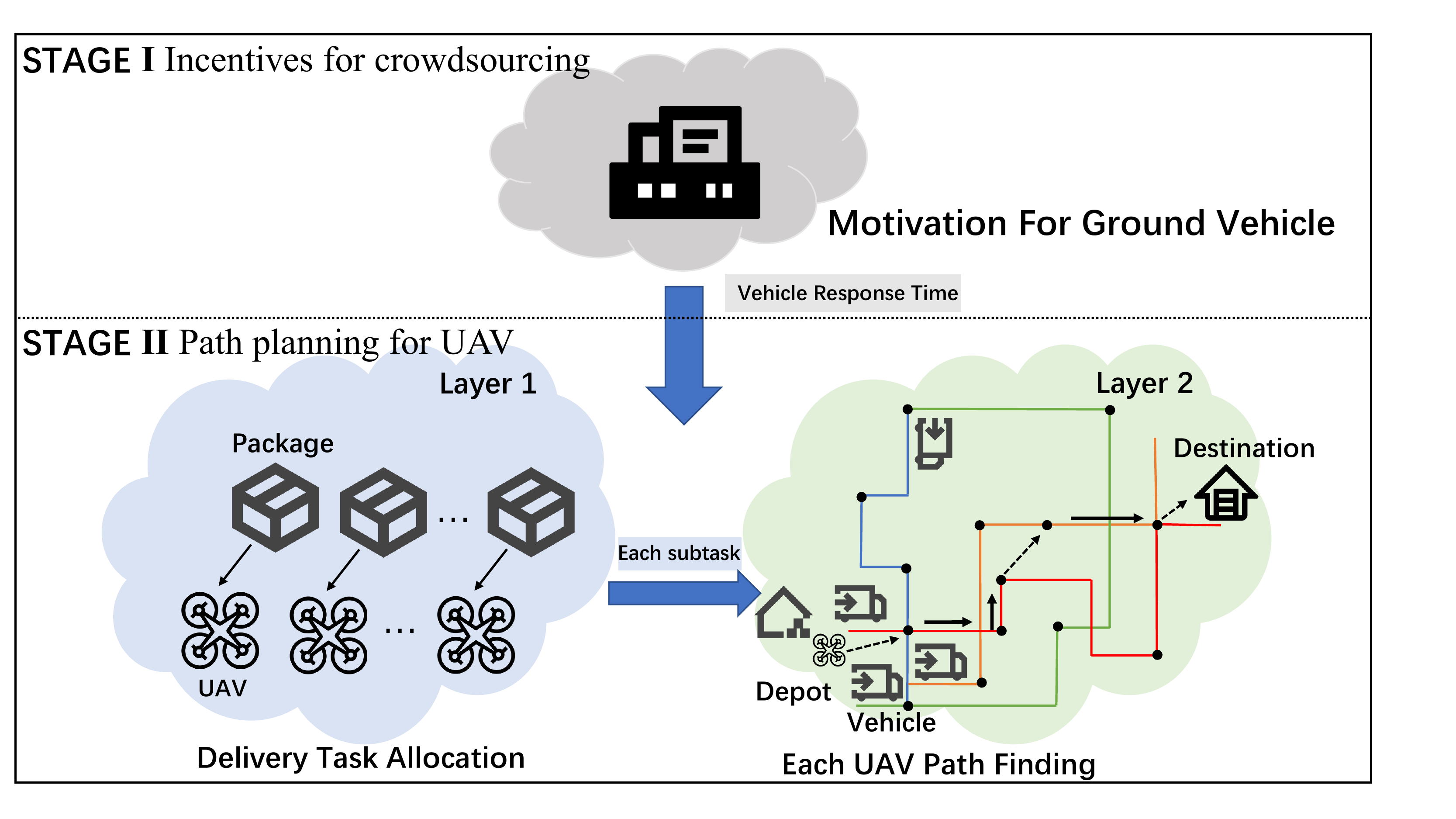}}
\caption{Two-stages multimodal logistics system: In stage I, the incentive mechanism for the crowdsourcing-based hitching services is designed, and the expected vehicle response time is predicted for the following path finding algorithm design. In stage II, the path planning problem is decomposed into two layers: first assign delivery tasks to all UAVs in Layer 1, and then find the shortest time-consuming path for each UAV on the basis of the vehicle response time in Layer 2. In the UAVs path finding section, we use the solid point to represent the interchange point, the dotted line to represent the flight route of UAVs and the solid line to represent the transit route of UAVs.}
\label{fig1}
\end{figure}

\subsection{Approach}
In a crowdsourcing-based air-ground collaborative logistics system, the behavioral decisions of the ground vehicles will affect the UAV 
transportation efficiency. If the UAV waits long time for the ground vehicles' response to assist in transportation, the passage time of the corresponding road section will be prolonged, which will lead to the delay of UAV mission completion time. Therefore, it is crucial to design an effective crowdsourcing incentive mechanism to encourage the ground vehicles to assist delivery. The passage time of each road section will also influence the design of subsequent task allocation and path planning algorithms. However, higher incentives can shorten the UAV waiting time, but cause an increase in UAV platform's cost. Therefore, we will design a dynamic pricing scheme to best balance the reward expenditure to ground vehicles and the expected vehicle response time of each road section.

Completing all package delivery requests means that each package needs to be assigned to one and only one UAV. So this problem can be viewed as a variation of the traveler problem. Because of the limitations of UAV carrying capacity and energy storage, we specify that a UAV can only deliver one package at a time and return to some depot for charging after completing the delivery task (charging time for UAVs is ignored here). The delivery time of a package by a UAV consists of the UAV flight time and the ride time including the vehicle response time (i.e., UAV waiting time) and the vehicle traveling time when carrying the UAV. We describe the energy storage limit of each UAV as a maximum flight time constraint, and each UAV must not exceed the maximum flight time when performing its delivery tasks. In addition, to avoid collisions between UAVs, multiple UAVs can't land at the same interchange point at the same time. The above joint optimization problem for UAV-vehicle path planning can be modeled as mixed integer linear programming. However, most of the existing mixed integer linear programming algorithms can only handle small-scale models, which are not suitable for large number of UAVs/packages and vast traffic network scenarios. We decompose the joint path planning problem into two layers of decision models for discussion, and design algorithms with low computational complexity at each layer, respectively.

We formulate our multimodal logistics system as two stages as shown in Fig. \ref{fig1}. In stage I, we predict the vehicle response time at each interchange point based on our designed dynamic pricing scheme, and update this information into the time weights of the interchange routes in the traffic network for the following path planning. In Stage II, we first analyze the task allocation of UAVs to solve the problem of which UAV delivers which packages and the order of delivery between packages, and then the path finding for each delivery subtask. To be specific, for task allocation in Layer 1, we temporarily ignore the specific route of the UAV from the depot to the package delivery address. We only decide which packages are sent from which depot and which UAV is responsible for which series of packages to be delivered, depending on the estimated time (the shortest path from a certain depot to the destination) between the depot and delivery location. We use linear programming algorithms with relaxation conditions that accomplish the task allocation problem in polynomial time. For path finding in Layer 2, we take one UAV to deliver a package from a depot and then return to a depot as a subtask, and perform route planning for the UAV fleet to execute the allocated delivery subtask. To avoid collisions between multiple UAVs, only a limited number of UAVs are allowed to stay at each interchange. We adopt the idea of CBS to design an efficient MAPF algorithm called conflict avoidance-based path search (CABPS), i.e., each UAV jointly maintains the same list of interchange occupancy information and performs its own path planning based on this constraint. 

\subsection{Framework And Workflow}

In Stage I, we motivate the ground vehicles to offer riding for UAVs by proper incentive mechanism design, e.g., monetary reward. However, the ground vehicles are mobile and their private costs for carrying the UAVs for certain distance is unknown. Under this circumstance, the waiting time for the UAVs to call for ground vehicles at different interchange points at different time is variable and unknown. In the following, we design a dynamic pricing scheme to balance the response time of ground vehicles and the monetary payment under the incomplete information about the ground vehicles' random arrivals at certain interchange point and private carrying costs. On the basis of the designed dynamic pricing scheme, the expected response time of ground vehicles at each interchange can be generated, which will be taken into consideration when performing path planning for each UAV. 



\begin{algorithm}[t]
\caption{UAVTaskAllocationPathPlanning.}
\begin{algorithmic}[1]
\REQUIRE Network Graph $G_T$; Location of depots $V_D$; 

        Location of packages $V_G$; Number of UAV $N$; 

        Waiting time for each interchange point $W$;
\ENSURE Paths of each UAV $P$;

\STATE Initiate constraints on interchange points $I_c$; Paths of each UAV $P$;
Each UAV delivery order $O\gets \text{TaskAllocation}(G_T,V_D,V_G,N)$; $i\gets 0$; Sum of delivered packages $s\gets 0$;
\WHILE{$s<|O|$}
\FOR{each UAV order $n=1:N$ \textbf{in} $O$}
\STATE $P_n^i\gets \text{CABPS}(G_T,O_n^i,W,I_c)$;
\STATE update constraints on interchange points $I_c$ based on current path of UAV;
\STATE $s\gets s+1$;
\ENDFOR
\STATE $i\gets i+1$;
\ENDWHILE
\RETURN $P$
\end{algorithmic}
\label{UTAPP}
\end{algorithm}

In Stage II, we propose a comprehensive algorithm UAV-Task-Allocation-Path-Planning for building an efficient air-ground cooperative multimodal logistics system. It guarantees a near-optimal solution in polynomial time. Our goal is to plan delivery paths for $N$ UAVs after accepting requests for $M$ packages.

The detailed design is given in \textbf{Algorithm 1}. Denote the traffic network graph as $G_T=(V_T,E_T)$, where the vertex set $V_T=V_G\cup V_D\cup V_I$, and each directed edge $(u,v)$ for every $u,v\in V_T$ contains a weight $w_{uv}$ denotes the passage time from $u$ to $v$ and an attribute $a_{uv}$ denotes the power consumption of the UAV from $u$ to $v$. We divide the directed edges $(u,v)$ into two types, one is the flight edge, which indicates that the UAV flies from $u$ to $v$. The other is the transit edge, which indicates that the UAV hitches a ground vehicle from $u$ to $v$. It is worth mentioning that if the directed edge $(u,v)$ is a transit edge, its power consumption $a_{uv}$ is $0$. Relatively, it has a longer passage time $w_{uv}$ (the passage time from $u$ to $v$ includes the UAV's waiting time and the carrying time from $u$ to $v$) than the flight edge.



\textbf{Layer 1 (line 1):} \emph{Task Allocation} is the first layer of our algorithm, which takes the traffic network graph $G_T$, the set of depot location $V_D$, the set of package locations $V_G$, and the number of UAVs $N$ as inputs, returns the set of UAV paths $O$, where $O_n^i$ indicates the $i^{th}$ delivery subtask for the $n^{th}$ UAV. Each subtask $O_n^i$ consists of three values: starting depot, package address and returning depot.

\textbf{Layer 2 (lines 2-10):} The second layer \emph{Multi-Agent-Path Finding} uses the Conflict Avoidance Based Path Search (CABPS) algorithm to calculate the path for each UAV separately. CABPS takes the $i^{th}$ subtask $O_n^i$ delivered by the $n^{th}$ UAV, the current waiting time estimate for each interchange and the occupancy time of each interchange as inputs, it gives the specific delivery path $P_n^i$ for each UAV's subtask $O_n^i$.

Our system eventually returns the set of delivery path $P$ for all UAVs. $P_n^i$ refers to the path planned by the $n^{th}$ UAV performing its own $i^{th}$ delivery subtask, and we use $\text{Time}({P_n})$ to indicate the total time taken by the $n^{th}$ UAV to complete all its delivery tasks. The purpose of our algorithm is to minimize ${\max}_{n\in N}\text{Time}({P_n})$.
Note that in the second layer of the algorithm, the paths for all delivery subtasks of the UAVs are not computed simultaneously. The delivery subtask of each UAV is computed successively once the previous subtask is accomplished, by considering the conflicts between UAVs. In the following sections, we will describe the design of the dynamic pricing mechanism and the two-layer algorithms in more detail.

\section{Stage I: Mechanism Design for Ground Vehicle Crowdsourcing}
\label{sec3}
In this section, we will discuss the incentive mechanism design for crowdsourcing at a typical interchange point. Consider a discrete time horizon with time slot $t=0,...,T_w$. When the UAV arrives at the interchange station, it sends the request and pricing reward to nearby vehicles. We use $s(t)=1$ to denote appropriate crowdsourcing vehicles' (such as trucks, buses, etc.) arrivals at certain interchange point at time $t$, and $s(t)=0$ otherwise. The probability of $s(t)=1$ can be viewed as the traffic density $\alpha$ (e.g., number of vehicles per meter, $veh/m$) at this interchange point. The vehicle decides whether to accept the current price $p(t)$ and provide hitching service according to its private cost $c\in[0,b]$ for carrying the UAVs, where the upper bound $b$ is estimated from historical data. In this paper, we consider non-trivial traffic density $\alpha$. Otherwise, the UAV platform should always set the price $p(t)$ to be the upperbound $b$ to encourage the rarely arriving ground vehicles to provide hitching services.

The vehicle response time $W(t)$ at time $t$ starting from the moment the UAV sent hitching requests (i.e., $t=0$) is affected by the traffic density $\alpha$ and the dynamic pricing $p(t)$. If a ground vehicle appears at time $t$ (i.e., $s(t)=1$) and its carrying cost is less than the offered price $p(t)$ (i.e., $c \leq p(t)$), the vehicle will provide hitching service. Otherwise, the vehicle response time $W(t+1)$ at time $t+1$ increases from $W(t)$ to $W(t)+1$. Therefore, starting from time slot $0$ with initial UAV waiting time $W(0)=0$, the dynamics of the vehicle response time is as follows:

\begin{equation}\label{eq1}
{W} (t+1)=\left\{\begin{aligned}
&{W}(t),\qquad & c \leq p(t) \& s(t)=1,\\
&{W}(t)+1, & \text { otherwise. }
\end{aligned}\right.
\end{equation}

According to the Cumulative Distribution Function $F(c)$ of the vehicles' private costs and the traffic density $\alpha$ at the interchange point, the probability of crowdsourcing vehicles accepting hitching service at time slot $t$ is $\alpha F(p(t))$. Therefore, at time slot $t+1$, the conditional expectation of the vehicle response time is:

\begin{equation}\label{eq2}
\begin{aligned}
\mathrm{E}[W(t+1) \mid W(t); p(t), \alpha] &=W(t) \alpha F(p(t))+(W(t)+1)\\
&\qquad(1-\alpha F(p(t))) \\
&=W(t)+1-\alpha F(p(t)).
\end{aligned}
\end{equation}
Denote $\bar{W}(t+1):=\mathrm{E}[W(t+1) \mid W(t); p(t), \alpha]$. Thus, the dynamics of the expected response time can be obtained as follows:
\begin{equation}\label{eq3}
\bar{W}(t+1)=\bar{W}(t)+1-\alpha F(p(t)).
\end{equation}

At each time slot $t$, the expected payment to the ground vehicle is $p(t) \alpha F(p(t))$. In order to maximize the benefits of the platform, the dynamic price $p (t)$ should not exceed the maximum private cost $b$ of the vehicle. Consider the uniform CDF of vehicles' private costs, i.e., $F(p(t))= p (t)/b$. The goal of the UAV platform is to design dynamic prices $p(t) \in[0, b]$, $t \in\{0, \ldots, T_{w}\}$ to minimize the summation of total expected payment and vehicle response time:

\begin{equation} \label{eq4}
U(T)=\min _{p(t) \in[0, b], t \in\{0, \ldots, T_w\}} \sum_{t=0}^{T_w} \rho^{t}\left(\bar{W}^{2}(t)+\frac{\alpha p^{2}(t)}{b}\right),
\end{equation}

\begin{equation} \label{eq5}
\text { s.t. }\bar{W}(t+1)=\bar{W}(t)+\left(1-\alpha \frac{p(t)}{b}\right).
\end{equation}
where $\rho \in(0,1)$ is the discount factor.

The above dynamic programming problem is not easy to solve by considering the huge number of price combinations over time. In the following, we will analyze the optimal solution to this dynamic problem by using dynamic control techniques.

\newtheorem{prop}{\bf Proposition}[section]
\begin{prop}\label{prop1}
The optimal dynamic pricing $p(t), t \in\{0, \ldots, T_{w}\}$ as the optimal solution to the dynamic program (\ref{eq4})-(\ref{eq5}) is monotonically increasing in $\bar{W}(t)$ and
given by
\end{prop} 

\begin{figure}[t]
\centerline{\includegraphics[width=0.4\textwidth]{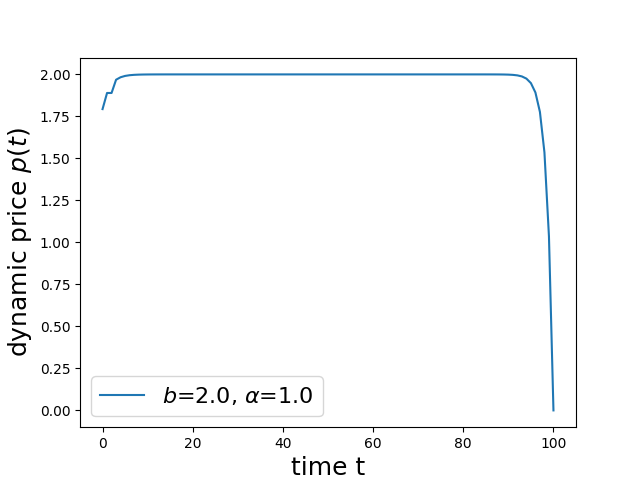}}
\caption{Optimal dynamic pricing $p(t)$ versus time $t$ with $\alpha=1$ over $T_w=100$ time slots.}
\label{fig2}
\end{figure}

\begin{figure}[t]
\centerline{\includegraphics[width=0.4\textwidth]{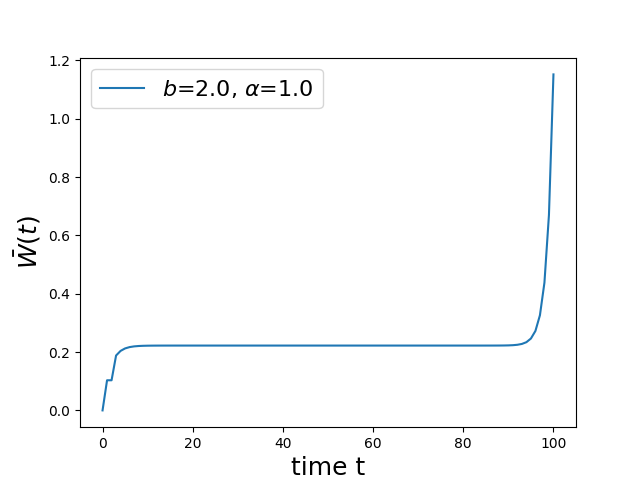}}
\caption{Vehicle response time (UAV waiting time) $\bar{W}(t)$ versus time $t$ with $\alpha=1$ over $T_w=100$ time slots.}
\label{fig3}
\end{figure}

\begin{equation}\label{eq6}
p(t)=\frac{\rho  M_{t+1}+2 \rho  Q_{t+1}(\bar{W}(t)+1)}{2+2 \rho  Q_{t+1} \frac{\alpha}{b}},    
\end{equation}
with $p(T_{w}) = 0$, and the vehicle response time $\bar{W}(t)$ at time $t, t\in \left [ 1,T_{w}\right ]$ is
\begin{equation}\label{eq7}
 \bar{W} (t)=
\frac{2-\rho M_{t} \frac{\alpha}{b}}{2+2 \rho Q_{t} \frac{\alpha}{b}} + \sum_{s=1}^{t-1} \frac{2-\rho M_{s} \frac{\alpha}{b}}{2+2 \rho Q_{s} \frac{\alpha}{b}} \prod_{i=s+1}^{t} \frac{1}{1+\rho Q_{i} \frac{\alpha}{b}},
\end{equation}
where
\begin{equation}\label{eq8}
   Q_{t}=1+\frac{\rho Q_{t+1}}{1+\rho Q_{t+1} \frac{\alpha}{b}},
\end{equation}
\begin{equation}\label{eq9}
    M_{t}=\frac{\rho\left(M_{t+1}+2 Q_{t+1}\right)}{1+\rho Q_{t+1} \frac{\alpha}{b}},
\end{equation}
with $\bar{W}(0)=0$, $Q_{T_w} = 1$, $M_{T_w} = 0 $ on the boundary.

\textbf{Proof}: Denote the cost objective function from initial time $t$ as

\begin{equation}\label{eq10}
 J(p, t)=\sum_{s=t}^{T_w} \rho^{s-t}\left(\bar{W}^{2}(s)+\frac{\alpha}{b} p^{2}(s)\right),  
\end{equation}
and the value function given the initial response time $\bar{W}(t)$ as

\begin{equation}\label{eq11}
V(\bar{W}(t), t)=\min _{\{p(s) \in[0, b]\}_{s=t}^{T_w}}(J(p, t) \mid \bar{W}(t)).
\end{equation}

Then, we have the dynamic programming equation at time~$t$:
\begin{equation}\label{eq12}
\begin{aligned}
  V(\bar{W}(t), t)= \min _{p(t) \in[0, b]} &\left (   \bar{W}^{2}(t)+\frac{\alpha}{b} p^{2}(t)\right ) \\
&+ \rho V(\bar{W}(t+1), t+1),
\end{aligned}
\end{equation}
\begin{equation}
\text { s.t. }\bar{W}(t+1)=\bar{W}(t)+\left(1-\alpha \frac{p(t)}{b}\right).\tag{5} 
\end{equation}

According to the first-order condition $ \frac{\partial V(\bar{W}(t), t)}{\partial p(t)}=0$  when solving (\ref{eq12}) in the backward induction process, we notice that  $p(t) $ is a linear function of  $\bar{W}(t) $. Thus, the value function in (\ref{eq12}) should follow the following quadratic structure with respect to $ \bar{W}(t) $ :
\begin{equation}\label{eq14}
   V(\bar{W}(t), t)=Q_{t} \bar{W}^{2}(t)+M_{t} \bar{W}(t)+S_{t},
\end{equation}
yet we still need to determine $Q_{t}, M_{t}, S_{t}$. This will be accomplished by finding the recursion in the following.

First, we have $Q_{T_w}=1, M_{T_w}=0, S_{T_w}=0 $ on the boundary due to $V(\bar{W}(T_w), T_w)=\bar{W}^{2}(T_w) $. Given  $V(\bar{W}(t+1), t+1)=   Q_{t+1} \bar{W}^{2}(t+1)+M_{t+1} \bar{W}(t+1)+S_{t+1}$  as in (\ref{eq14}), the dynamic programming equation at time $t$ is
\begin{equation}\label{eq15}
  \begin{aligned}
V(\bar{W}(t), t) 
=& \min _{p(t)}\left(\bar{W}^{2}(t)+\frac{\alpha}{b} p^{2}(t)+\rho Q_{t+1} \bar{W}^{2}(t+1)\right.\\
&\left.+\rho M_{t+1} \bar{W}(t+1)+\rho S_{t+1}\right) .
\end{aligned}  
\end{equation}

Substitute $ \bar{W}(t+1) $ in (\ref{eq5}) into (\ref{eq15}) and let $\frac{\partial V(\bar{W}(t), t)}{\partial p(t)}=0 $, we obtain the optimal price $ p(t)$  in (\ref{eq6}). Then, we substitute  $p(t)$  in (\ref{eq6}) into $ V(\bar{W}(t), t) $ in (\ref{eq15}), and obtain $V(\bar{W}(t), t) $ as a function of  $Q_{t+1}, M_{t+1}, S_{t+1}$  and  $\bar{W}(t) $. Finally, by reformulating  $V(\bar{W}(t), t) $ in (\ref{eq15}) and noting that  $V(\bar{W}(t), t)=   Q_{t} \bar{W}^{2}(t)+M_{t} \bar{W}(t)+S_{t}$ , we obtain the recursive functions of $ Q_{t}$  and $ M_{t} $ in (\ref{eq8}) and (\ref{eq9}). Substitute   $p(t)$ in (\ref{eq6}) into (\ref{eq5}), we obtain the expected age $\bar{W}(t)$   in (\ref{eq7}).$\hfill\blacksquare$

Figs. \ref{fig2} and \ref{fig3} simulate the system dynamics over time under the optimal dynamic price $ p(t)$  in (\ref{eq6}). We can see that $ p(t)$  is low initially, allowing a modest increase in the response time to save the UAV platform's costs. The price  $p(t)$  then increases with the increased response time to encourage the ground vehicles' help until both of them reach steady states. This is consistent with the monotonically increasing relationship between $ p(t)$  and $\bar{W}(t) $ in (\ref{eq6}). However, when approaching the end of the time horizon  $T_w=100$, the price $ p(t)$  drops to $0$  without worrying about its effect on the future waiting time. As a result, the vehicle response time  $\bar{W}(t) $ increases again but lasts only a few time slots.

\newtheorem{lemma}{\bf Lemma}[section]
\begin{lemma} \label{lemma1}
As $T_w \rightarrow \infty$, $Q_{t}$ in (\ref{eq8}) and $M_{t}$ in (\ref{eq9}) respectively converge to the following steady-states:
\end{lemma}

\begin{equation}\label{eq16}
    Q=\frac{1}{2}\left(1-\frac{b(1-\rho)}{\rho \alpha}\right.\\
\left.+\sqrt{\left(1-\frac{b(1-\rho)}{\rho \alpha}\right)^{2}+\frac{4 b}{\rho \alpha}}\right),
\end{equation}

\begin{equation}\label{eq17}
    M=\frac{2 \rho Q}{1-\rho+\rho Q \frac{\alpha}{b}}.
\end{equation}

\textbf{Proof}: Starting from the boundary condition $ Q_{T_w}=1, M_{T_w}=0 $, according to  $Q_{t}=1+\frac{\rho Q_{t+1}}{1+\rho Q_{t+1} \frac{\alpha}{b}}$  in (\ref{eq8}) and  $M_{t}=  \frac{\rho\left(M_{t+1}+2 Q_{t+1}\right)}{1+\rho Q_{t+1} \frac{\alpha}{b}} $ in (\ref{eq9}), we can obtain $ Q_{t} $ and   $M_{t}$ backward given $ Q_{t+1}^{b} $ and $ M_{t+1}$  at next time slot. Rewrite $ Q_{t}$  above as
\begin{equation} \label{eq18}
  Q_{t}=1+\frac{\rho}{\frac{1}{Q_{t+1}}+\rho \frac{\alpha}{b}}.  
\end{equation}

Starting from $ Q_{T_w}=1$  and note that  $\frac{\rho}{\frac{1}{Q_{t+1}}+\rho \frac{\alpha}{b}}>0 $, we have  $Q_{T_w-1}>1>Q_{T_w}$. Note that (\ref{eq18}) is increasing with $ Q_{t+1}  $ and  $\lim _{Q_{t+1} \rightarrow \infty}=1+\frac{b}{\alpha}<+\infty$ , we can conclude that  $\left\{Q_{T_w}, Q_{T_w-1}, \ldots, Q_{0}\right\}  $ is bounded increasing sequence in the reverse time order and will converge to the steady-state $Q$ , which can be solved as (\ref{eq16}) by removing the time subscripts from (\ref{eq8}).

By removing the time subscripts from (\ref{eq9}), the stead-ystate  $M $ is solved as  $M=\frac{2 \rho Q}{1-\rho+\rho Q \frac{\alpha}{b}} $ in (\ref{eq17}). In the following, we will show that  $\left\{M_{T_w}, M_{T_w-1}^{b}, \ldots, M_{0}\right\} $ is bounded increasing sequence in the reverse time order and converges to  $M$. According to  $M_{t}=\frac{\rho\left(M_{t+1}+2 Q_{t+1}\right)}{1+\rho Q_{t+1} \frac{\alpha}{b}} $ in (\ref{eq9}), define $ y\left(Q_{t+1}\right) $ as the right-hand side of (\ref{eq9}) given the steady-state $M$, i.e.,
\begin{equation}\label{eq19}
 y\left(Q_{t+1}\right)=\frac{\rho\left(M+2 Q_{t+1}\right)}{1+\rho Q_{t+1} \frac{\alpha}{b}} .   
\end{equation}

\begin{figure}[t]
\centerline{\includegraphics[width=0.4\textwidth]{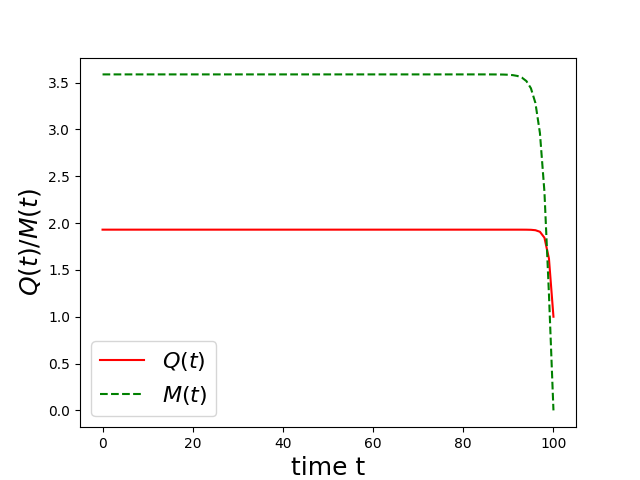}}
\caption{$Q_{t}$ in (\ref{fig8}) and $M_{t}$ in (\ref{eq9}) versus time $t$ with $T_w = 100$ time slots, $b=2$,$\alpha$=1,$\rho$=0.9.}
\label{fig4}
\end{figure}

Take the derivative of  $y\left(Q_{t+1}\right) $ with respect to  $Q_{t+1} $, we have
\begin{equation}\label{eq20}
    \frac{d y\left(Q_{t+1}\right)}{d Q_{t+1}}=\frac{2 \rho-M \rho^{2} \frac{\alpha}{b}}{\left(1+\rho Q_{t+1} \frac{\alpha}{b}\right)^{2}}.
\end{equation}
According to $ M=\frac{2 \rho Q}{1-\rho+\rho Q \frac{\alpha}{b}} $, we have  $M \rho \frac{\alpha}{b}=   \frac{2 \rho Q \rho \frac{\alpha}{b}}{1-\rho+\rho Q \frac{\alpha}{b}}=\frac{2 \rho}{\frac{1-\rho}{Q \rho \frac{\alpha}{b}}+1}<2 $. Thus,$  \frac{d y\left(Q_{t+1}\right)}{d Q_{t+1}}>0 $ and $ y\left(Q_{t+1}\right) $ increases with  $Q_{t+1} $. Since $\left\{Q_{T_w}, Q_{T_w-1}, \ldots, Q_{0}\right\} $ is an increasing sequence and converges to  $Q$, $y\left(Q_{T_w}\right)<   y\left(Q_{T_w-1}\right)<\cdots<y(Q) $.

Starting from  $M_{T_w}=0 $, we have  $M_{T_w-1}=\frac{2 \rho}{1+\rho \frac{\alpha}{b}}>   0=M_{T_w} $. Since $ \frac{2 \rho Q_{t+1}}{1+\rho Q_{t+1} \frac{\alpha}{b}} $ increases with $ Q_{t+1}$,  the equation set  $\left\{M_{t}=\frac{\rho\left(M_{t+1}+2 Q_{t+1}\right)}{1+\rho Q_{t+1} \frac{\alpha}{b}}\right\} $ intersects with the vertical axis with  $\frac{2 \rho Q_{T_w-1}}{1+\rho Q_{T_w-1} \frac{\alpha}{b}} \leq   \frac{2 \rho Q_{T_w-2}}{1+\rho Q_{T_w-2} \frac{\alpha}{b}} \leq \cdots \leq \frac{2 \rho Q_{T_w-4}}{1+\rho Q_{T_w-4} \frac{\alpha}{b}}$. The slope of each equation satisfies  $0<\frac{\rho }{1+\rho Q_{t+1} \frac{\alpha}{b}}<1 $. Therefore, starting from $ M_{T_w-1}>0 $, we have $ M_{T_w-2}>M_{T_w-1} $ based on $ M_{T_w-2}=\frac{\rho\left(M_{T_w-1}+2 Q_{T_w-1}\right)}{1+\rho Q_{T_w-1} \frac{\alpha}{b}} $. Then, from  $M_{T_w-2}$, we have  $M_{T_w-3}>M_{T_w-2}  $ based on $ M_{T_w-3}=\frac{\rho\left(M_{T_w-3}+2 Q_{T_w-3}\right)}{1+\rho Q_{T_w-3} \frac{\alpha}{b}}$. Thus, we can obtain$  M_{T_w}<M_{T_w-1}<M_{T_w-2}<M_{T_w-3}<\cdots  $until $M_{t}=M_{t+1}=M$.$\hfill\blacksquare$

Fig. \ref{fig4} simulates the dynamics of $Q_{t}$ in (\ref{eq8}) and $M_{t}$ in (\ref{eq9}). We can see that both $Q_{t}$ and $M_{t}$ fast converge to
their steady-states in a few rounds. In the following, we will show the steady-state of the dynamic pricing $p(t)$ in (\ref{eq6}) for infinite time horizon case.

\newtheorem{rem}{\bf Remark}[section]
\begin{rem}\label{prop1}
For the infinite time horizon case, the
optimal dynamic pricing is simplified from (\ref{eq6}) to

\begin{equation} \label{eq21}
p^{\infty}(t)=\frac{\rho M+2 \rho Q\left(\bar{W}^{\infty}(t)+1\right)}{2+2 \rho Q \frac{\alpha}{b}},
\end{equation}
and the vehicle response time $\bar{W}^{\infty}(t)$ at time $t$ is
\begin{equation}\label{eq22}
\bar{W}^{\infty}(t)=\frac{2-\rho M \frac{\alpha}{b}}{2+2 \rho Q \frac{\alpha}{b}} \frac{1-\left(\frac{1}{1+\rho Q \frac{\alpha}{b}}\right)^{t}}{1-\frac{1}{1+\rho Q \frac{\alpha}{b}}}.
\end{equation}

As $t \rightarrow \infty$, according to $Q$ in (\ref{eq16}) and $M$ in (\ref{eq17}), and note that
$\frac{1}{1+\rho Q \frac{\alpha}{b}}<1$, the vehicle response time converge to the steady-state $\bar{W}$:
\begin{equation}\label{eq23}
\bar{W}=\lim _{t \rightarrow \infty} \bar{W}^{\infty}(t)=\frac{(1-\rho)\left(1+\rho Q \frac{\alpha}{b}\right)}{\rho Q\left(\frac{\alpha}{b}(1-\rho)+\rho Q\left(\frac{\alpha}{b}\right)^{2}\right)},
\end{equation}

and the optimal dynamic pricing converges to
\begin{equation}\label{eq24}
\lim _{t \rightarrow \infty} p^{\infty}(t)=\frac{b}{\alpha}.
\end{equation}
which shows that $p^{\infty}(t)\leq b$ always holds when $\alpha\geq 1$.
\end{rem} 


\section{Stage II: Layer 1: Task Allocation}
\label{sec4}

\subsection{Algorithm Overview}
This section focuses on the first layer of the stage II. We design the UAV task allocation scheme to traverse all packages and minimize the maximum delivery time of each UAV. Package nodes $V_G$ and the depot nodes $V_D$ in the traffic network graph $G_T$ are picked out to generate the task allocation graph, which is denoted as $G_A=(V_A,E_A)$, where the set of vertices $V_A=V_G\cup V_D$, and the weight $w_{uv}$ of directed edge $(u,v)\in E_A$ is the predicted passage time from position $u$ to position $v$. We run the shortest path search algorithm in the traffic network graph $G_T$ to obtain the predicted passage time $w_{uv}$ between all points of the task allocation graph $G_A$. We solve such a problem in $G_A$ to find a shortest path to access all packages, which means we find the delivery sequence of all packages and it has the smallest total predicted passage time.

This layer aims to design the task allocation scheme $P_n=d_1g_1d_2g_2\dots d_kg_kd_{k+1},n\in\{1,\dots ,N\}$ to minimize the longest UAV task completion time, i.e., $\min {\max}_{n\in N}\text{Time}(P_n)$, where $\text{Time}(P_n)$ is the total time for the $n^{th}$ UAV to complete the corresponding package delivery service and return to depot $d_{k+1}$ under task allocation $P_n$ from depot $d_1$ in delivery order $\{g_1,\dots ,g_k\}$. Task allocation $P_n$ contains the information of which UAV delivers which packages and the delivery order. To find the task allocation scheme for $N$ UAVs, we first consider the minimum connection tours (MCT) problem that connects all packages:
\begin{equation}
    P1: \min{\sum}_{(u,v)\in E_A}w_{uv}x_{uv}, \label{P1:1}
\end{equation}
\begin{equation}
\begin{aligned}
    s.t. {\sum}_{(g,d)\in E_A}x_{gd}=&{\sum}_{(d,g)\in E_A}x_{dg}=1, \\ &\forall d\in V_D,g\in V_G, \label{P1:2}
\end{aligned}
\end{equation}
\begin{equation}
    {\sum}_{(v,d)\in E_A}x_{vd}={\sum}_{(d,v)\in E_A}x_{dv}=1, \forall d\in V_D, \label{P1:3}
\end{equation}
\begin{equation}
    x_{dg},x_{gd}\in \{0,1\}, \forall d\in V_D,g\in V_G, \label{P1:4}
\end{equation}
\begin{equation}
    x_{dd^{'}}\in {\mathbb{N}}_{\ge 0}, \forall d,d^{'}\in V_D, \label{P1:5}
\end{equation}
where constraint (\ref{P1:2}) indicates that each package address $g\in V_G$ must be visited once and comes from one depot, and the UAV must return to one depot after completing delivery. Constraint (\ref{P1:3}) indicates that the UAVs entering and exiting each depot are equal, which paves the way for subsequently transforming this problem into a minimum cost circulation problem and designing algorithms in polynomial time. Constraint (\ref{P1:4}) indicates that the edge connected to the package address is used at most once. The last constraint (\ref{P1:5}) indicates that multiple round trips between depots are possible.


\begin{algorithm}[t]
\caption{TaskAllocation.}
\begin{algorithmic}[1]
\REQUIRE Network Graph $G_T$; Location of depots $V_D$; 

        Location of packages $V_G$; Number of UAV $N$; 

\ENSURE Each UAV packages delivery order $O$;

\STATE Allocation graph $G_A\gets \text{SubGraph}(G_T,V_D,V_G)$;
\STATE Edge selection vector $X\gets \text{MCC}(G_A)$;
\STATE Connected components $C\gets \text{GetComponents}(G_A, X)$;
\WHILE{$|C|>1$}
\FOR{$c,c^{'}$\textbf{in }$C$}
\IF{$w_{dd^{'}}+w_{d^{'}d}$is minimum that $d\in c,d^{'}\in c^{'}$}
\STATE $C\gets \text{MergeComponent}(C,c,c^{'})$;
\ENDIF
\ENDFOR
\ENDWHILE
\STATE Get the tour that start from one depot then visit all the goods finally end at a depot $T\gets \text{GetTours}(C)$;
\STATE Split Tour $T$ into $N$ orders $O\gets \text{SplitTour}(T,N)$;
\RETURN $O$
\end{algorithmic}
\label{TA}
\end{algorithm}

The detailed design is given in \textbf{Algorithm 2}, in which we aim to minimize the sum of the estimated passage time between depots and packages while ensuring that each package is dispatched by one and only one depot.

\textbf{SubGraph}. At the task allocation layer, we seek a set of $N$ paths that minimize the maximum delivery time of any UAV. Here we ignore the specific routes from the depot to the package and only consider how they are connected. A new task allocation graph $G_A$ is generated from the network $G_T$ based on the known geographic locations of depots and packages, $G_A$ contains only the nodes of depots and packages, and the weights between them are the estimated passage times between the two locations. 

\textbf{MinimumCostCirculation (MCC)}.
The problem $P1$ can be viewed as an $N$ minimal visiting paths problem\cite{b12} and is NP-hard. Note that the structure of the minimal visiting paths problem is very similar to the MCC problem, and the latter can be solved by a polynomial-time algorithm\cite{b19}. In problem $P1$, $x_{uv}$ indicates the number of times to select the edge $(u,v)$, which is an integer. To reduce the computational complexity of the linear programming problem, we transform the integer constraints of the combinatorial optimization problem $P1$ into real numbers, thus transforming the problem into a MCC problem for analytical solution. Finally, we divide the minimum connected path obtained by linear programming into $N$ paths for each UAV.


\textbf{GetStrongConnectedComponent}. The solution $X=\{x_{uv}$, $u$, $v\in V_A\}$ given by MCC shows the number of times the edge $(u,v)$ is used in the task allocation graph $G_A$. The depot nodes are connected to the package nodes according to the edges used and forming multiple connected components. Then we combine all the connected components to generate the entire connected graph, which has edges covering the trips we need to traverse all the packages with the least cost.

\textbf{SplitTours}. After obtaining the trip that traverses all the packages with minimum cost, we distribute it evenly among all the UAVs. Evenly means that the estimated time of package delivery undertaken by each UAV is close to the total estimated time divided by the total number of UAVs $N$. The estimated package delivery time is not equivalent to the actual UAV delivery time. The time spent by the UAV includes the flight time, carrying time (hitching time), waiting time (vehicle response time), and the time spent by the UAV for conflict resolution. 

\subsection{Computational complexity and optimality}
Since we relax the variable $X=\{{x_{uv},u,v\in V_A}\}$ in linear programming from an integer constraint to a real number constraint, the MCT problem is converted to a MCC problem. For the MCC problem, we can compute an integer optimal solution in polynomial time\cite{b19,b20}.

Our goal is to minimize the maximum delivery time of UAVs, by returning the set $P={P_1,...,P_N}$, where $P_n$ indicates the sequence of package delivered by the $n^{th}$ UAV. Then our objective can be expressed as
\begin{equation}
    \text{minimize }{\max}_{n\in N}\text{Time}(P_n).
\end{equation}

Let the optimal task allocation set be $\widetilde{P}=\{\widetilde{P_1},\dots ,\widetilde{P_N}\}$, where $OPT:={\max}_{n\in N}\text{Time}(\widetilde{P_n})$. In the task allocation algorithm, linear programming usually returns more than one connected component, and we aim to find the minimum weights of the edges between depots to connect different connected components, which finally generates a strongly connected path over the graph. Such a path $T$ traverses all the packages from a depot and then back to a depot, and contains the newly added edges (the depot to another depot), which makes our result slightly deviate from the optimal solution. We observe that
\begin{equation}
\begin{aligned}
    |T|&\leq {\sum}_{n\in N}\text{Time}(\widetilde{P_n})+\alpha \mathop{\max}_{d,d^{'}\in V_D}(w_{dd^{'}}+w_{d^{'}d})
     \\ &\leq N\mathop{\max}_{n\in N}\text{Time}(\widetilde{P_n})+\alpha \mathop{\max}_{d,d^{'}\in V_D}(w_{dd^{'}}+w_{d^{'}d}),
\end{aligned}
\end{equation}
where $|T|$ is the total weights of the paths $T$ and $\alpha$ is the number of times to merge the connected components. We assign a task to each UAV by specifying that its package delivery prediction time is less than the total prediction time divided by total number of UAVs $N$. The worst outcome of this is that a UAV is assigned an additional task after it has been assigned exactly the average prediction time, such that we can derive that
\begin{equation}
    {\max}_{n\in N}\text{Time}(P_n)\leq |T|/N+\mathop{\max}_{d,d^{'}\in V_D,g\in V_G}(w_{dg}+w_{gd^{'}}).
\end{equation}

We find that the predicted time undertaken by any one UAV in this layer of the algorithm does not exceed the average predicted time plus the estimated time for one maximum package delivery time. Combining the above equations, we finally get the gap between our results and the optimal solution:
\begin{equation}
    {\max}_{n\in N}\text{Time}(P_n)\leq OPT+\beta +\gamma,
\end{equation}
where $\beta :=\frac{a}{N}\mathop{\max}_{d,d^{'}\in V_D}(w_{dd^{'}}+w_{d^{'}d})$, $\gamma :=\mathop{\max}_{d,d^{'}\in V_D,g\in V_G}(w_{dg}+w_{gd^{'}})$. Obviously, as the number $N$ of UAVs and the number $M$ of package delivery requests rises, our result gets closer to the optimal solution and solves in polynomial time.

\section{Stage II: Layer 2: Multi-Agent Path Finding}
\label{sec5}


\begin{figure*}[!t]
\centering
\subfloat[]{\includegraphics[width=2in]{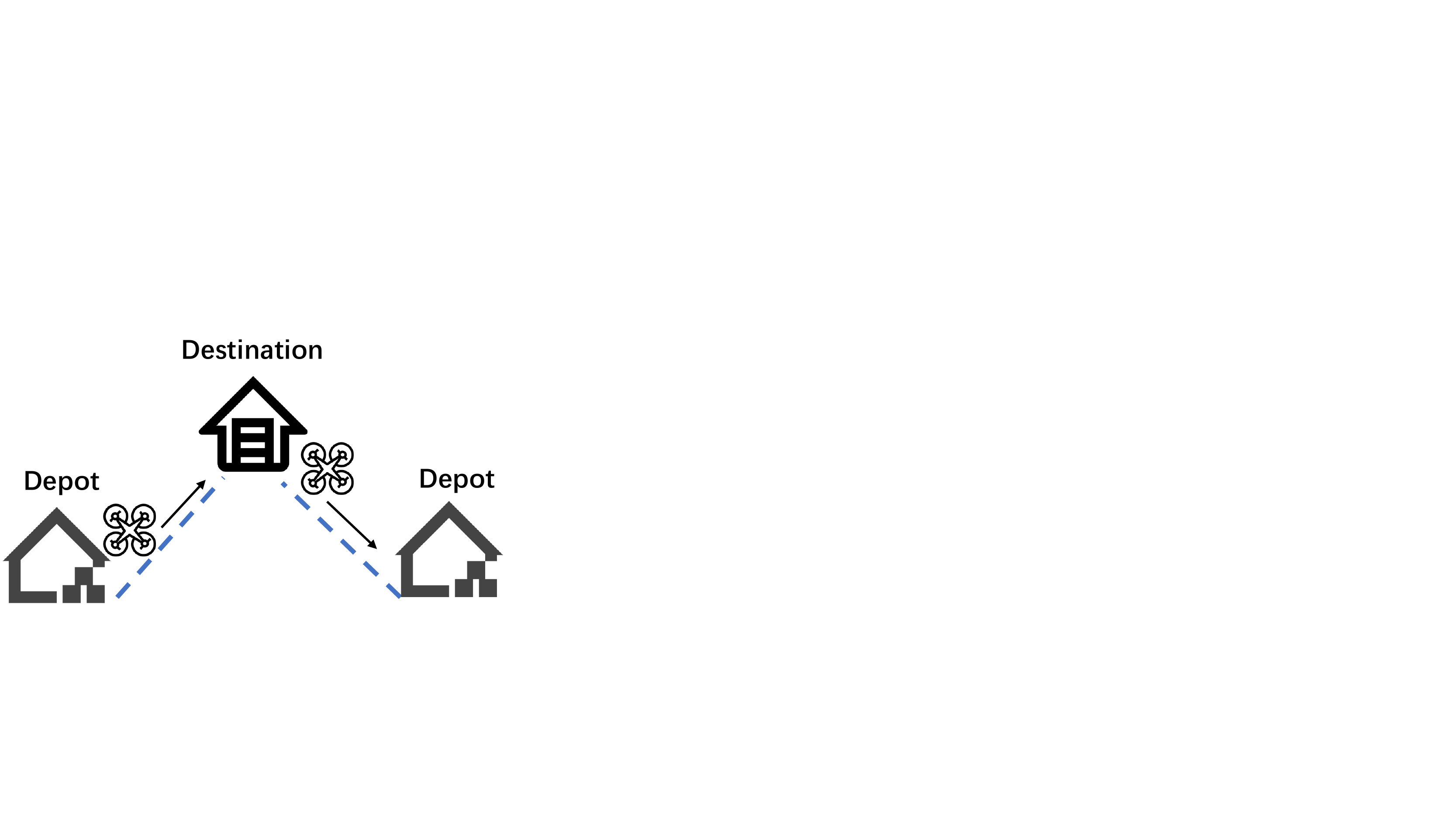}%
\label{flight}}
\hfil
\subfloat[]{\includegraphics[width=2in]{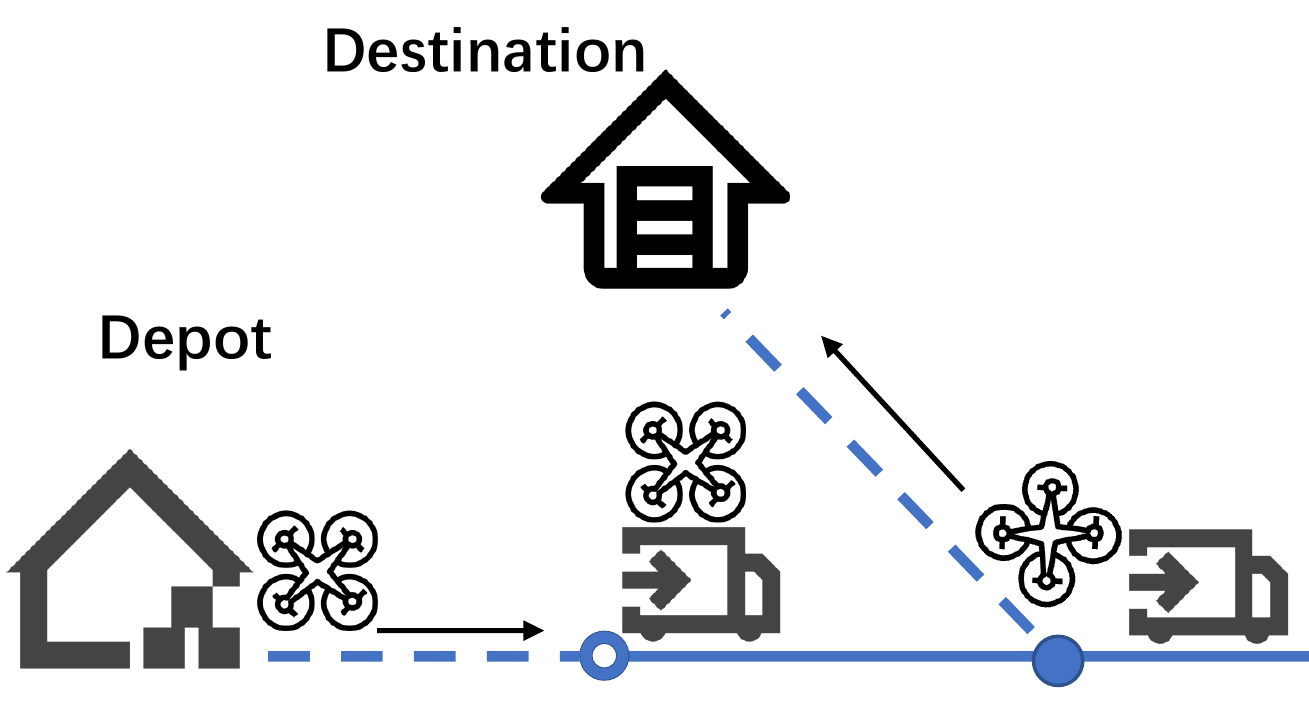}%
\label{singlehop}}
\hfil
\subfloat[]{\includegraphics[width=2in]{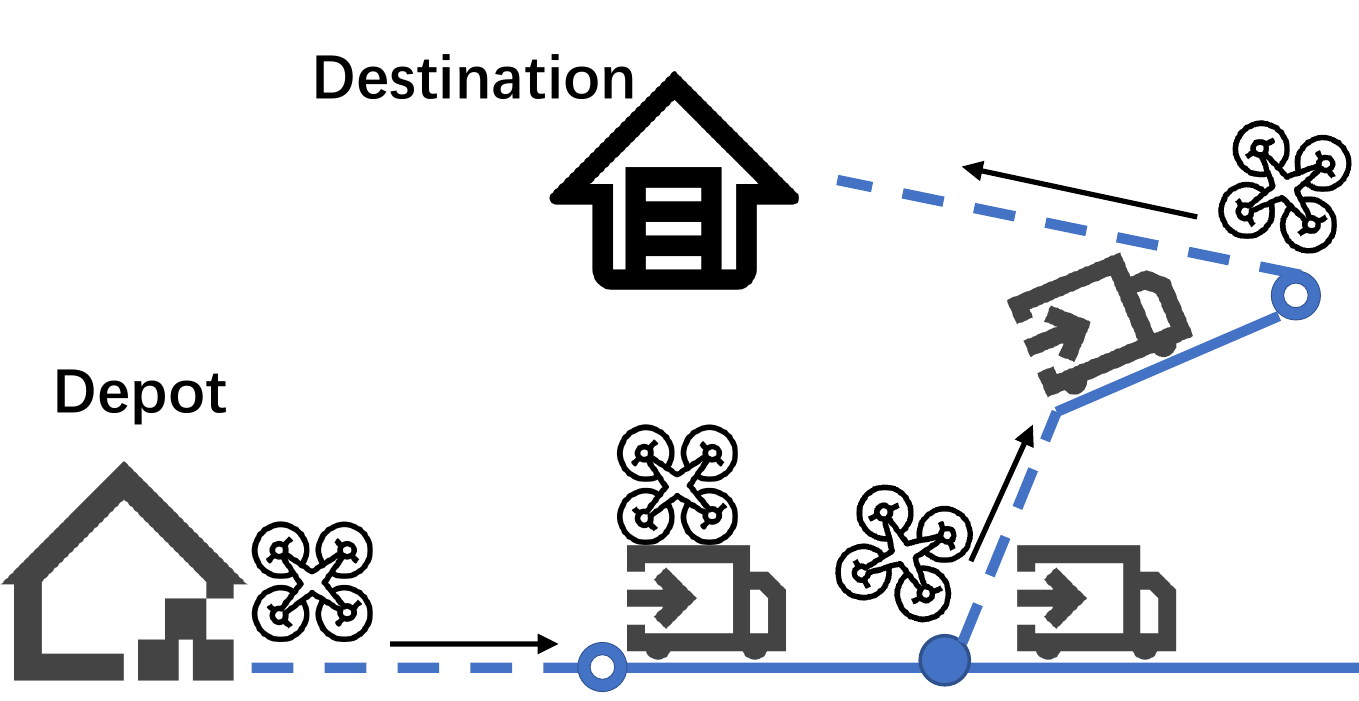}%
\label{multihop}}
\caption{Three types of UAV delivery networks, where the solid line represents the hitching route and the dotted line represents the flight route. (a) The UAV will not carry ground vehicles during delivery mission. (b) The UAV can only carry a ground vehicle once during a delivery mission. (c) The UAV can carry ground vehicles more than once during a delivery mission.}
\label{fig5}
\end{figure*}

According to the task allocation results of the UAVs in the first layer, the task of each UAV can be decomposed into a series of $dgd^{'}$ delivery subtasks, i.e., delivering packages from depot $d$ to destination $g$ and back to the same or different depot $d^{'}$. The conflict between UAVs comes from the space limitation of the ground vehicles and interchange points. Here, we define as only one UAV can stay at an interchange point at the same time to wait for ground vehicles.\footnote{Our algorithms can be extended to the case with more than one UAV staying at an interchange point at the same time, and the simulation result is shown in Section \ref{sec6}.} Based on whether the UAVs hitch on ground vehicles and whether hop between different ground vehicles for each subtask, we discuss the following three types of UAV delivery networks as shown in Fig.~\ref{fig5} (a) direct flight network (b) single-hop hitch network (c) multi-hop hitch network.

\subsection{Direct Flight Network}
In a direct flight network, UAVs are utilized for delivering packages without the help of ground vehicles. In this case, we consider that the UAVs deliver packages by flying from the depot to the package address directly, and thus the delivery time of a UAV is equal to the straight-line distance from the depot to the package address divided by the UAV's average flight speed. Though delivering directly by UAVs saves a lot of time especially during peak hours or in an a complex geographical environment, the energy limitation of the UAVs has restricted their service range. In the direct flight network, for each package delivery request executed by the UAV, we need to consider the constraint that whether the flight distance consumed in the whole journey exceeds its own limit. When the UAV cannot reach the designated location, we consider the delivery mission to fail. The multimodal logistics system can solve the flight distance limitation of direct flight network, and we compare the mission delivery failure rates of the three networks in Section \ref{sec6} later.

\subsection{Single-Hop Hitch Network}
\begin{algorithm}[t]
\caption{Single-Hop Path Finding.}
\begin{algorithmic}[1]
\REQUIRE Departure location $d$; 
        End point location $e$; 
        Transit sections set $P_r$;
\ENSURE Path of the current subtask $P$;
\STATE Initialize maximum aircraft time for UAVs $\overline{T}$; Minimum passage time $t_{min}\gets +\infty$;
\FOR{r \textbf{in} $P_r$}
\STATE $(i,j)\gets r$;
\IF{$t^{'}_{di}+t^{'}_{je}\leq \frac{\overline{T}}{2}$ \textbf{and} not exceeding the capacity of the interchange point $i$}
\IF{$t^{'}_{di}+t_{ij}+t^{'}_{je}\leq t_{min}$}
\STATE $t_{min}\gets t^{'}_{di}+t_{ij}+t^{'}_{je}$;
\STATE $P\gets \{d,i,j,e\}$;
\ENDIF
\ENDIF
\ENDFOR
\RETURN $P$
\end{algorithmic}
\label{TA}
\end{algorithm}
In a single-hop hitch network, the UAV hitch on a ground vehicle only once during the journey. For a delivery subtask of $dgd^{'}$, the UAV can depart from the depot $d$, fly to the interchange point $i$ to wait and hitch on the vehicle through the transit section $r=(i,j)$, and then fly from $j$ to the delivery destination $g$. After completing the subtask, it returns to the depot $d^{'}$ in order to perform the next subtask. Denote the passage time of the UAV from the depot to the package address be $T_{n,r,dg}=t^{'}_{di}+t_{ij}+t^{'}_{jg}$, where $t^{'}_{di}$ and $t^{'}_{jg}$ are the corresponding UAV flight time (which can be obtained from the UAV speed) and $t_{ij}$ is the ride time. Based on the crowdsourcing incentive mechanism, the vehicle response time for the transit section $r=(i,j)$ is predicted to be $\overline{W}_{i}$. So the ride time is $t_{ij}=\overline{W}_{i}+t^v_{ij}$, where $t^v_{ij}$ is the vehicle passage time for section $r$ (which can be obtained from the average vehicle speed). We decompose one delivery subtask into two segments $dg$ and $gd^{'}$. The transit sections $r$ and $r^{'}$ can be the same or different sections. The transit sections are directed edges and the set of all transit sections is $P_r$. With the objective of minimizing delivery subtask completion time, the UAV hitch problem is modeled as
\begin{equation}
\begin{aligned}
    P2:\min{\sum}_{r\in P_r}(T_{n,r,dg}y_{n,r,dg}+T_{n,r,gd^{'}}y_{n,r,gd^{'}}),
    \\ \forall n\in \{1,\dots ,N\}, \label{P2:1}
\end{aligned}
\end{equation}
\begin{equation}
    s.t. {\sum}_{n\in \{1,\dots ,N\}}(y_{n,r,dg}+y_{n,r,dg^{'}})\leq C_r,\forall r\in P_r, \label{P2:2}
\end{equation}
\begin{equation}
\begin{aligned}
    {\sum}_{r\in P_r}y_{n,r,dg}={\sum}_{r\in P_r}y_{n,r,gd^{'}}=1,
    \\ \forall n\in \{1,\dots ,N\}, \label{P2:3}
\end{aligned}
\end{equation}
\begin{equation}
\begin{aligned}
    {\sum}_{r\in P_r}({t^{'}_{n,r,dg}}y_{n,r,dg}+{t^{'}_{n,r,gd^{'}}}y_{n,r,gd^{'}})\leq \overline{T},
    \\ \forall n\in \{1,\dots ,N\}, \label{P2:4}
\end{aligned}
\end{equation}
\begin{equation}
    y_{n,r,dg},y_{n,r,gd^{'}}\in \{0,1\},\forall r\in P_r,\forall n\in \{1,\dots ,N\}, \label{P2:5}
\end{equation}
where constraint (\ref{P2:2}) indicates that the UAV that selects a transit section $r$ cannot exceed its capacity $C_r$ (no more than $C_r$ UAVs can stay at an interchange point at the same time), thereby avoiding UAV conflicts. The constraint (\ref{P2:3}) indicates that only one transit section can be selected from $d$ to $g$ and from $g$ to $d^{'}$. And constraint (\ref{P2:4}) indicates that the total flight time of the UAV cannot exceed its maximum flight time $\overline{T}$, where ${t^{'}_{n,r,dg}}$ is the flight time of the UAV $n$ that selects section~$r$.

A delivery subtask of $dgd^{'}$ can be divided into two parts: from $d$ to $g$ and from $g$ to $d^{'}$. Thus each UAV needs to invoke the path finding algorithm twice with half the energy constraint each time to perform a subtask and the detail of single-hop path finding is given in \textbf{Algorithm 3}. The algorithm takes the departure location $d$, destination $e$ and interchange sections $P_r$ as inputs. Single-Hop Path Finding algorithm only has to traverse all the interchanges once to find the path that takes the shortest time and does not exceed the UAV's flight limit.

\subsection{Multi-Hop Hitch Network}
In multi-hop hitch networks, UAVs can extend their delivery range by hopping between more than one ground vehicles in one delivery task, which makes the path finding problem more challenging due to more combinations of transit sections. When performing path planning, UAVs need to consider their own flight distance limitations and the conflicts between UAVs. CBS can solve the conflict problem between UAVs. It first ignores the conflict between UAVs in the upper layer and finds the optimal path for each UAV, then detects whether there is a conflict between the paths of each UAV in the lower layer. If a conflict occurs, CBS adds a new restriction and runs the path planning algorithm in the upper layer again. However, multiple agents operate on the same traffic network, and whenever a conflict is detected between UAVs, a new conflict restriction needs to be added for each conflicting UAV and a new path is planned for each UAV again. As the number of UAVs $N$ and the number of package delivery requests $M$ increases, the number of conflicts between UAVs will become larger and larger. Thus, the system will frequently plan new routes for UAVs, leading to degradation of network performance \cite{b17}. 

\begin{algorithm}[t]
\caption{CABPS.}
\begin{algorithmic}[1]
\REQUIRE Network Graph $G_T$; Departure location $d$; End point location $e$; Vehicle response time for each interchange point $\overline{W}$; Constraints on interchange points $I_c$;

\ENSURE Path of the current subtask $P$;
\STATE Initialize maximum aircraft time for UAVs $\overline{T}$; Initialize an empty Priority Queue $Q$;
\STATE Generate a new subgraph $G_S\gets \text{SubGraph}(G_T,d,e)$;
\STATE Push $d$ into Priority Queue $Q$;
\WHILE{$|Q|>0$}
\STATE current node $c\gets \text{pop}(Q)$;
\IF{$c==e$}
\RETURN Path $P$ from departure point to $c$;
\ENDIF
\FOR{$i$ \textbf{in} neighbors of point $c$}
\IF{$w_{ci}+$ total flight time cost from $d$ to $c<\frac{\overline{T}}{2}$}
\IF{edge$(c,i)$ \textbf{in} interchange edges of $G_S$}
\STATE Consider response times $\overline{W}$ and occupancy conflicts $I_c$ at interchange point;
\ENDIF
\STATE push $i$ into Priority Queue $Q$;
\ENDIF
\ENDFOR
\ENDWHILE
\RETURN empty path $P$

\end{algorithmic}
\label{CABPS}
\end{algorithm}

To deal with the performance degradation of CBS, we propose a conflict avoidance-based path search algorithm for the path optimization problem under multi-hop hitch network, which can be solved in polynomial time. Firstly, we search for the shortest path for UAVs by $\text{A}^{*}$ heuristic algorithm. When a UAV determines a clear path, we want to make sure other UAVs don't clash with it. So the system will updates the information about the occupied interchange points in the network as it plans the path for each UAV, and that occupancy information needs to be taken into account when planning paths for other UAVs. Thus we are able to ensure that no conflicts occur in the network. The detail of CABPS is given in \textbf{Algorithm 4}. The inputs to the algorithm are the entire network graph $G_T$, the departure location $d$, destination point $e$, the vehicle response time $\overline{W}$ of each interchange predicted based on the crowdsourcing incentive mechanism and the information about the current timestamp $I_c$ of each interchange being occupied.

\textbf{Step 1 (lines 1-2):} Construct a new path search graph $G_S=(V_S,E_S)$ from the original graph based on the departure node $d$ and end node $e$ of the current subtask, where $V_S=d\cup e\cup V_I$. We connect such edges like $(u,v)\in E_S$ that $u,v\in V_S$, for any edge $(u,v)$ with two properties, one is the passage time spent from $u$ to $v$, and the other is the UAV flight time consumed. We define the path search graph $G_S$ as follows.

\newtheorem{defi}{\bf Definition}[section]
\begin{defi}  There is only one depot node and one package node in the path search graph $G_S$, and the rest are interchange points. For an edge $(u,v)$, its properties are divided into the following cases: (i) If one of $u$ and $v$ is a depot or package node, the passage time of the edge is the straight-line distance from $u$ to $v$ divided by the average speed of the UAV, and the consumption time of the edge is the flight time of the UAV. (ii) If both $u$ and $v$ are interchange points and the edge $(u,v)$ is a transit route, the passage time of the edge is the distance of the journey from $u$ to $v$ divided by the average speed of the vehicle, and the consumption of flight time is $0$. (iii) Both $u$ and $v$ are interchange points, but the edge $(u,v)$ is not a transit route. The passage time of this edge is the straight-line distance from $u$ to $v$ divided by the UAV speed and the consumption time of the edge is the flight time of the UAV.
\end{defi}

\textbf{Step 2 (lines 3-17):} Use the $\text{A}^{*}$ heuristic algorithm\cite{b23} to search for the shortest path in the subgraph $G_S$. $\text{A}^{*}$ uses a priority queue to store the information of historical paths from start node to the candidate nodes, and the heuristic function is the shortest path search algorithm from the current node to the destination node. In the path search process, we save the paths that do not consume more flight than the limit, and the passage time consumed by the current path includes the delivery time, waiting time at the interchange and the additional waiting time due to conflict avoidance.

\textbf{Step 3 (lines 18):} If $\text{A}^{*}$ cannot find a path that matches the condition, we return an empty set to indicate that the delivery subtask failed.

\subsection{Feasibility and Time Complexity}

For the direct flight network, the UAV only considers the straight-line distance from the origin location to the destination point, so its result is unique. As for the single-hop hitch network where the UAV only carries the ground vehicle once during the process from the delivery task, the path combination of the single-hop hitch network is limited and only related to the number of interchange routes in the network, thus we mainly discuss the multi-hop hitch network here.

\subsubsection{The shortest path in time} 
$\text{A}^{*}$ selects a new node when performing an iterative search based on the cost of the path and an estimate of the cost required to extend the path to the destination. $\text{A}^{*}$ selects the minimized path as
\begin{equation}
    f(n)=g(n)+h(n),
\end{equation}
where $n$ is the candidate nodes on the path, $g(n)$ is the passage time spent from the starting point to $n$. $h(n)$ is a heuristic function to estimate the cost of the shortest path from $n$ to the destination, which we represent by the time-consuming shortest path between two points calculated by Dijkstra's algorithm. In the UAV path search, the actual resulting shortest path must not take less time than the unconstrained shortest path, considering the energy storage limitation of UAV, thus our heuristic function design is reasonable.

When $\text{A}^{*}$ stops searching, it indicates that it has found a path from the starting point to the goal that has an actual cost lower than the estimated cost of any other path. 
Thus
$\text{A}^{*}$ will never overestimate the actual cost of reaching the destination, and is guaranteed to return the lowest cost path from the starting point to the destination.

\subsubsection{Avoiding conflicting paths}
In CABPS, we ensure that each planned path will not be affected by the subsequent path. That is, each time a path is planned for a UAV, we no longer change it, but update the occupancy information of this path on the interchange points of the network for the subsequent path planning. The occupancy information is a time-stamped interval, its lower bound is the arrival time of the UAV at the site, and upper bound is the arrival time plus the response time of the vehicles on the ground. The situation is considered as a conflict if the UAV's arrival time is within the interval of the interchange's occupancy timestamp. Different from CBS in which the conflicting paths are jointly planned, in CABPS, when a conflict occurs, the original path will not be replanned, and the subsequent UAV needs to wait until another UAV leaves the interchange point, which means that more waiting time is introduced. 
In the following lemma, we show that even though CABPS may get longer delivery time by reducing the computational burden, the performance gap between CBS and CABPS is small. 

\begin{lemma} $quasiOPT$ is the approximate optimal solution of $cbsOPT$, where $cbsOPT$ and $quasiOPT$ are the passage time of the paths obtained by CBS and CABPS, respectively.
\end{lemma}


\textbf{Proof}: Suppose there are two paths $p$, $q$ with conflicting in certain time stamp. Denote $T_r$ as the time spent on path $r$. When a conflict occurs between paths $p$ and $q$, the problem is divided into two subproblems according to the definition of CBS\cite{b17}, i.e., $p$ maintains the original path and $q$ adds restrictions to find a new path $q^{'}$ or $q$ maintains the original path and $p$ adds restrictions. Then the extra time spent on the path after CBS resolution can be expressed as $\min(T_{q^{'}}-T_q,T_{p^{'}}-T_p)$. According to the CABPS definition, we always guarantee that a path is not subject to change. Now assume that the invariable path is $p$. The occupancy time of path $p$ at the interchange point that conflicts with path $q$ is $\overline{W}_p$, and the extra time can be expressed as $\min(T_{q^{'}}-T_q,\overline{W}_p)$. When the extra time is $\overline{W}_p$, it indicates that we maintain the original path $q$, and the extra time for path $q$ is the waiting time for the UAV on path $p$ leaving the occupied interchange point, i.e., $\overline{W}_p$. Thus, we have

\begin{equation}
    cbsOPT=T_p + T_q + \min(T_{q^{'}}-T_q, T_{p^{'}}-T_p),
\end{equation}
\begin{equation}
    quasiOPT=T_p + T_q + \min(T_{q^{'}}-T_q,\overline{W}_p).
\end{equation}

The different between CBS and CABPS is the extra time spent on conflict resolution and it can be observed that
\begin{equation}
\begin{aligned}
    quasiOPT=&cbsOPT+\min(T_{q^{'}}-T_q,\overline{W}_p)
    \\ &-\min(T_{q^{'}}-T_q, T_{p^{'}}-T_p).
\end{aligned}
\end{equation}
Assuming that $T_{q^{'}}<T_q$, then $q^{'}$ will be the optimal path before the conflict occurs, which causes the contradiction. Similarly $T_{p^{'}}>T_p$. So we have $quasiOPT <= cbsOPT + \min(T_{q^{'}}-T_q, \overline{W}_p)$. We note that $quasiOPT <= cbsOPT + \overline{W}_p$ holds constantly, where $\overline{W}_p$ is the vehicle response time depending on the incentive mechanism and traffic density.

Now we extend the problem to $N$ paths in conflict. According to the definition of CBS, the conflicts of the first two paths are resolved first and branching is done based on their conflicts. Denote the time spent to resolve the conflicts of $N$ paths as ${\sum}_{i\in \{1,\dots ,N-1\}}{cbsOPT}_i$. The CABPS algorithm determines each path in order, so the time spent to resolve the conflicts of $N$ paths is ${\sum}_{i\in \{1,\dots ,N-1\}}{quasiOPT}_i$. In the worst case, we have
\begin{equation}
\begin{aligned}
    &{\sum}_{i\in \{1,\dots ,N-1\}}{quasiOPT}_i
    \\ &\leq {\sum}_{i\in \{1,\dots ,N-1\}}{cbsOPT}_i+(N-1){\max}_{r\in P_r}\overline{W}_r
    \\ &\leq (1+C){\sum}_{i\in \{1,\dots ,N-1\}}{cbsOPT}_i,
\end{aligned}
\end{equation}
where $C=\frac{(N-1){\max}_{r\in P_r}\overline{W}_r}{{\sum}_{i\in \{1,\dots ,N-1\}}{cbsOPT}_i}$ and $P_r$ is the set of conflicting paths. According to the dynamic pricing, the maximum vehicle response time can be predicted. For small vehicle response time, the factor $C$ is small, which makes the bound on $quasiOPT$ closer to the optimal solution $cbsOPT$.$\hfill\blacksquare$


\begin{table}[t]
\renewcommand\arraystretch{1.5}
    \centering
    \caption{The average computation time (seconds) of the task allocation algorithm on a fixed network $G_T=(V_T,E_T)$, where $|V_T|=846$, $M$ is the number of packages and $K$ is the number of depots.}
    \begin{tabular}{c c c c c c c c}
    \hline
        M & K=2 & K=3 & K=4 & K=5 & K=10 & K=15 & K=20 \\ \hline
        10 & 0.010 & 0.016 & 0.037 & 0.088 & 0.258 & 0.923 & 1.875 \\ 
        20 & 0.013 & 0.019 & 0.041 & 0.100 & 0.294 & 0.975 & 2.192 \\ 
        50 & 0.013 & 0.024 & 0.048 & 0.133 & 0.350 & 1.182 & 2.646 \\ 
        100 & 0.015 & 0.021 & 0.051 & 0.128 & 0.360 & 1.203 & 2.685 \\ 
        200 & 0.019 & 0.034 & 0.082 & 0.216 & 0.533 & 1.802 & 3.790 \\ 
        400 & 0.029 & 0.047 & 0.123 & 0.263 & 0.735 & 2.644 & 5.024 \\ 
        600 & 0.036 & 0.057 & 0.152 & 0.403 & 11.585 & 2.880 & 5.989 \\ \hline
    \end{tabular}
    \label{table1}
\end{table}

\subsubsection{Computational complexity}
As mentioned above, in CABPS, the planned UAV route will not be changed, and later UAVs can only choose to replan route to avoid conflict or wait for the previous UAV to leave. Thus, the shortest path search algorithm only needs to be run once for each UAV.
Noting that the UAV cannot exceed its own maximum flight distance during delivery, but adding this restriction to the shortest path problem makes it NP-hard. Motivated by \cite{b24}, we extend the $\text{A}^{*}$ into MultiConstraint Shortest Path ($\text{A}^{*}\_ \text{MCSP}$) algorithm and use it as a low-level search algorithm for CABPS. As $\text{A}^{*}\_ \text{MCSP}$ prunes the unnecessary search space by forward-looking information and eligibility tests, it significantly reduces the computational complexity and solve the problem in polynomial time. Thus, the time complexity of CABPS is also polynomial.


\section{Simulations}
\label{sec6}

\subsection{Task Allocation}

In this section, we validate our results via simulations. The traffic network is generated on the basis of some real-world scenarios. 

In Section \ref{sec4}, we show that our method can obtain an near-optimal solution. In Table \ref{table1}, we verify the efficiency of the task allocation algorithm for different numbers of depots and package delivery requests. Our approach is to randomly select some nodes on the network as depots or packages, and get an evaluation matrix with size of $(K+M)\times (K+M)$, where $K$ is the number of depots and $M$ is the number of packages. Then we run the task allocation algorithm on this matrix. As shown in Table \ref{table1}, the task allocation running time of our algorithm grows polynomially as $K$ and $M$ increase.

\subsection{Multi-Agent Path Finding}

In this sub-section, we verify the efficiency of our results in MAPF layer, by examining the success rate of UAV deliveries, impacts of the interchange capacities, the number of UAVs and depots on the package delivery time, respectively.

\begin{figure}[t]
\centerline{\includegraphics[width=0.45\textwidth]{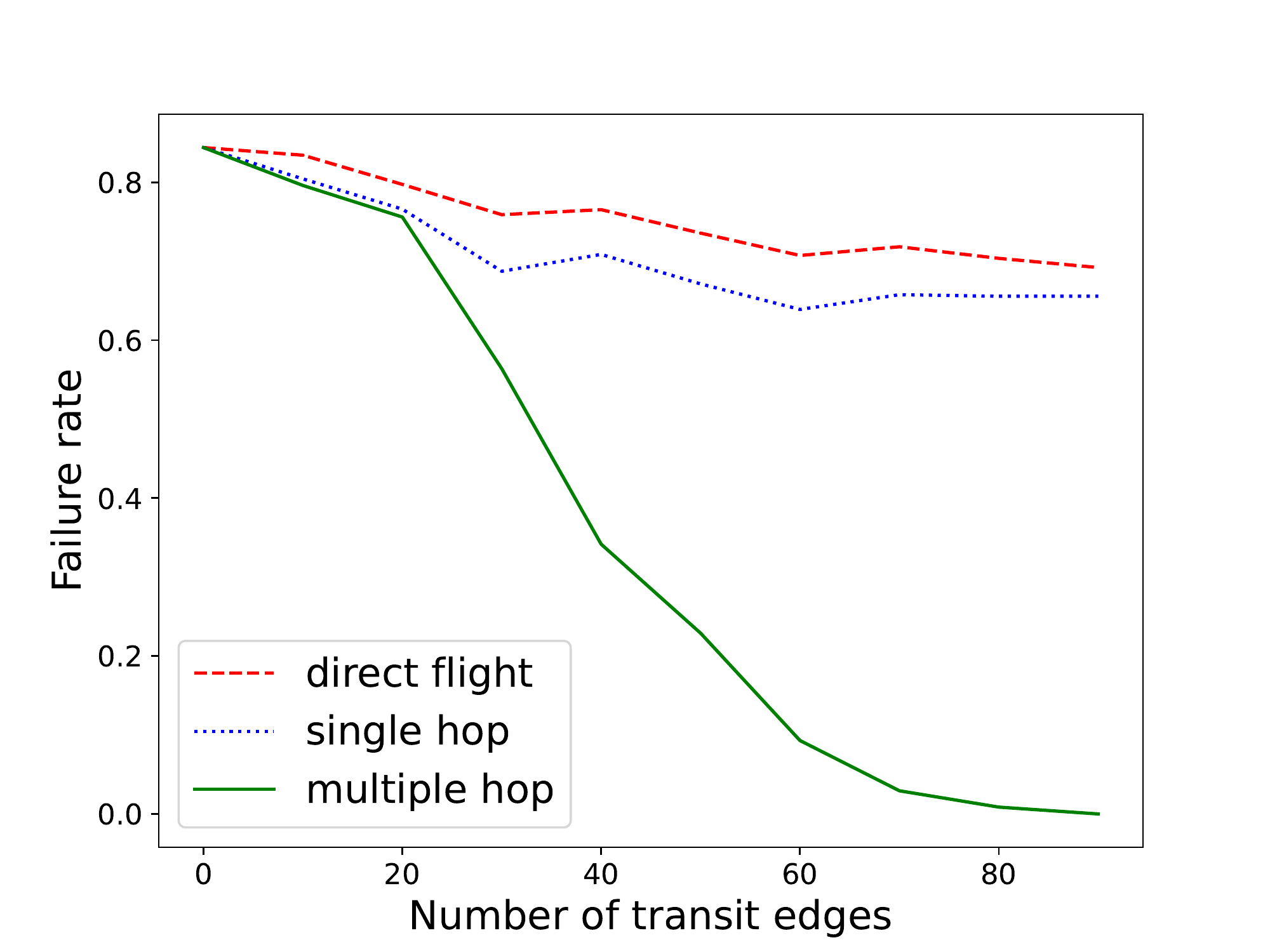}}
\caption{The average delivery failure rate of three UAV delivery networks vs. number of transit edges.}
\label{fig6}
\end{figure}

\begin{figure}[t]
\centerline{\includegraphics[width=0.45\textwidth]{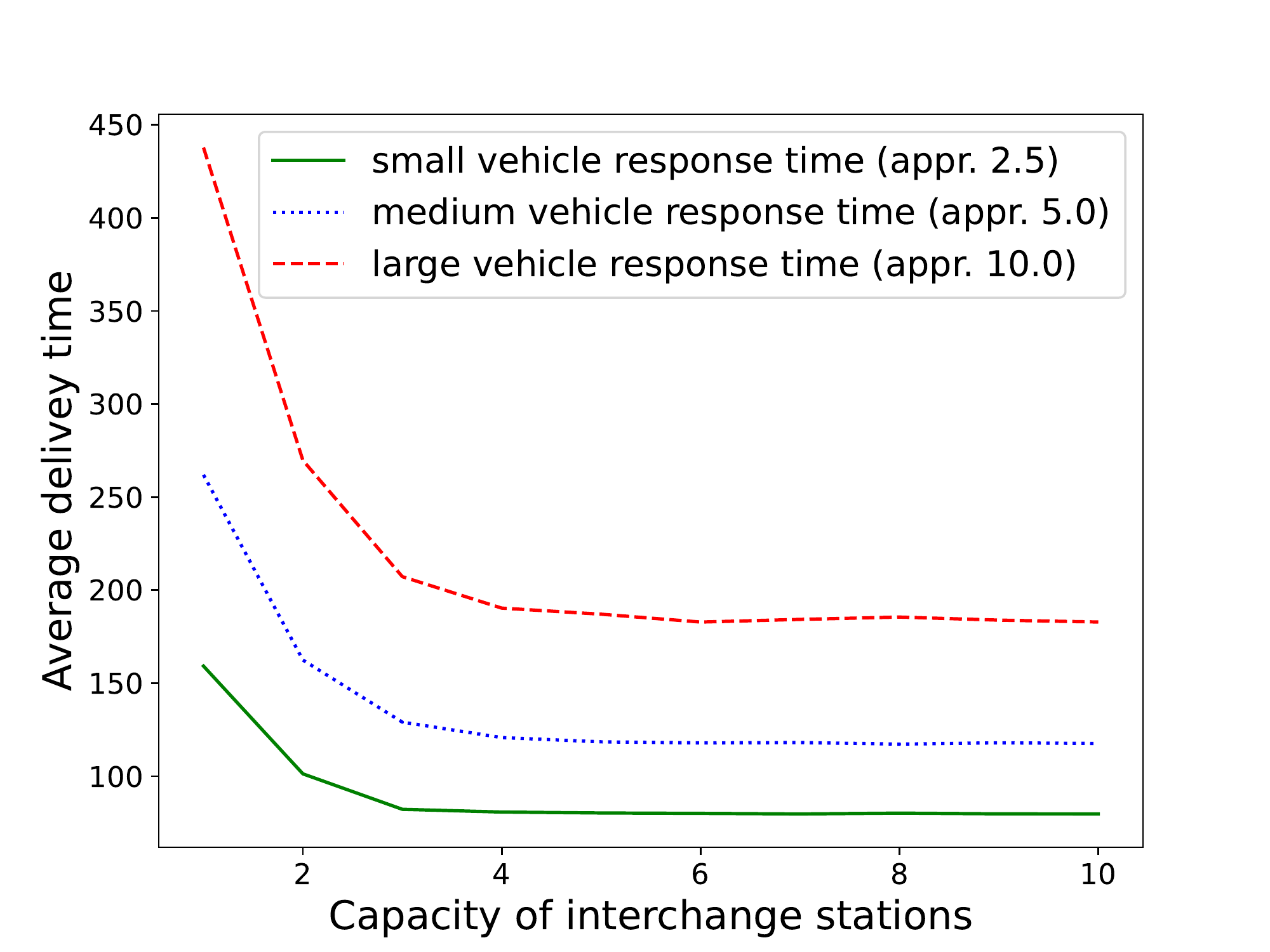}}
\caption{The average delivery time vs. capacity of interchange point for different maximum vehicle response time. }
\label{fig7}
\end{figure}

Even though all the packages are allocated to certain UAVs at the task allocation layer, there are still some packages that the UAVs cannot deliver due to the energy constraint. If the CABPS algorithm cannot find a feasible path or its execution time exceeds $300$ seconds, we consider this package delivery task a failure. Fig. \ref{fig6} compares the failure rates of UAVs delivering package under three types of UAV delivery networks. We use the number of transit edges in the networks as variables, and $50$ experiments are conducted for each number of transit edges. The network is randomly generated for each experiment, and we take the average of the final results. To demonstrate the effectiveness of our multi-hop hitch network in improving the range of UAV deliveries, we set the maximum flight distance of the UAV to be small. As shown in Fig. \ref{fig6}, the average package delivery failure rate for the multi-hop hitch network eventually drops to $0\%$, comparing to $70\%$ and $65\%$ for the direct flight and single-hop hitch networks. It also shows that the performance of our multi-hop hitch network improves significantly as the number of interchange edges increased.

\begin{figure}[t]
\centerline{\includegraphics[width=0.5\textwidth]{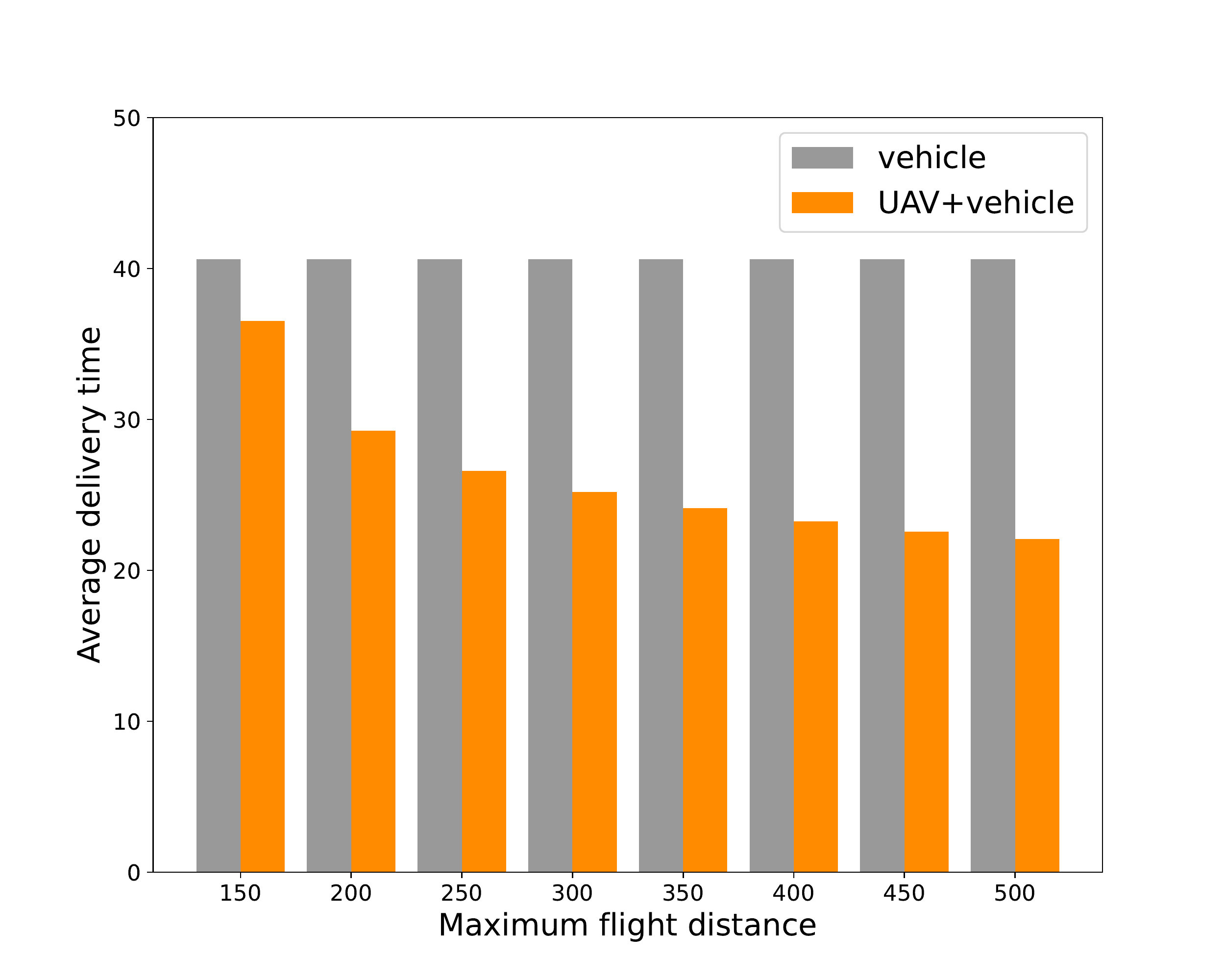}}
\caption{Average delivery time vs. maximum flight distance for vehicles only and our multimodal logistics system.}
\label{fig8}
\end{figure}

We extend to the case with more than one UAV staying at an interchange point at the same time, which reduces the probability of conflict in the logistics system. Assuming that the capacity of each interchange point is $\bar{C}$ which means that up to $\bar{C}$ UAVs can stay at an interchange point simultaneously. Fig. \ref{fig7} shows how the UAV's delivery time changes with the capacity of the interchange point for different vehicle response time. For each setting, $50$ experiments are conducted. It shows that the delivery time reduces for smaller vehicle response time with less additional cost associated with conflict avoidance by UAVs. In addition, as the capacity of the interchange points increases, the delivery time decreases due to less conflicts.

We compare the average delivery time of vehicles only and our multimodal logistics system. Considering the speed limit of the vehicle in the urban area, we set the speed of the UAV to $1.3$ times the speed of the vehicle. The delivery time of the vehicle is obtained based on the shortest path. We use the flight distance limit of the UAV as an experimental variable and set the number of interchange paths to $60$ in order to ensure that the UAV can deliver most of the packages. $50$ experiments are conducted separately for different flight limits. It is shown in Fig. \ref{fig8} that, as the UAV's maximum flight distance increase, the delivery time for the air-ground combination delivery method reduces. This is because, within large flight distance, the UAV relies less on the ground vehicles to complete the package delivery task, and thus takes less time.


\begin{table*}[ht]
\renewcommand\arraystretch{1.5}
    \centering
    \caption{The average time (in seconds) for sub-task path planning calculation time, sub-task delivery time and maximum delivery time for any UAV to complete a delivery task, where $N$ is the number of UAVs and $K=|V_D|$ is the number of depots. $50$ experiments are conducted for each setting.}
    \begin{tabular}{c c c c c c c}
    \hline
        \multicolumn{6}{c}{\{ Avg Calculation Time, Avg Delivery Time, Max Delivery Time\}} \\ \hline
        N &  K=1 & K=3 & K=5 & K=10 & K=20 & K=30 \\ \hline
        1 & \{0.237, 52.0, 5202\} & \{0.037, 27.8, 2779\} & \{0.005, 15.4, 1541\} & \{0.005, 12.9, 1289\} & \{0.003, 9.5, 945\} & \{0.0002, 7.5, 747\} \\ 
        5 & \{0.236, 52.1, 1244\} & \{0.038, 27.8, 666\} & \{0.005, 15.3, 404\} & \{0.005, 12.7, 312\} & \{0.003, 9.2, 220\} & \{0.0001, 7.0, 170\} \\ 
        10 & \{0.243, 52.2, 688\} & \{0.038, 27.8, 383\} & \{0.004, 15.2, 224\} & \{0.005, 12.6, 182\} & \{0.002, 8.6, 128\} & \{0.0002, 6.5, 96\} \\ 
        20 & \{0.244, 52.3, 392\} & \{0.038, 27.7, 223\} & \{0.004, 15.0, 139\} & \{0.005, 12.3, 112\} & \{0.002, 8.3, 78\} & \{0.0001, 6.0, 60\} \\ 
        30 & \{0.245, 52.5, 301\} & \{0.039, 27.7, 179\} & \{0.004, 14.8, 108\} & \{0.005, 12.1, 94\} & \{0.002, 8.1, 64\} & \{0.0001, 5.8, 50\} \\ \hline
    \end{tabular}
    \label{table2}
\end{table*}

Finally, we verify the impact of the number of depots and UAVs on the network performance. The experimental results for different configurations are given in Table \ref{table2}. As the number of depots rises, the UAV can be allocated to the depots closer to it for delivery, leading to shorter delivery time. The increase in the number of UAVs allows us to distribute tasks out evenly, which reduces the time to complete all package delivery requests. 

\section{Conclusion}\label{sec_conclusion}
We propose an effective air-ground multimodal logistics system based on crowdsourcing. To deal with the UAVs' energy storage limitation problem, the ground vehicles are invited to provide hitching services for UAVs. We formulate the crowdsourcing-based multimodal logistics problem by two-stages. In Stage I, a dynamic pricing scheme is proposed to motivate the ground vehicles to help UAV
delivery, which balance the monetary reward and waiting time for UAVs for cost minimization. In Stage II, the path planning over the traffic network is discussed. We divide the problem into two layers to solve. In the first layer, we use an approximate optimal solution algorithm to distribute delivery tasks in polynomial time. In the second layer, a multi-UAV path search algorithm is proposed based on conflict avoidance, which is an approximate solution of CBS and maintains its performance in the case of a large number of conflicts. Finally, several simulations are conducted to verify that our approach is efficient. It is shown that our system substantially increases the delivery range of the UAV (its delivery success rate increased by $325\%$ in our experiments) and its delivery time is much less than vehicle deliveries.

%

\bibliographystyle{IEEEtran}
\bibliography{refs}

\end{document}